\documentstyle[mnras_cite,graphicx,pifont,amssymb,amsmath]{mn}

\newcommand{\ti}{\mathrel{\times}}
\newcommand{\y}{\hbox{$\surd$}}
\newcommand{\w}{\hbox{$\omega$}}

\title[The nuclei of a volume-limited sample of galaxies I]
{Multiwavelength study of the nuclei of a volume-limited sample of galaxies I: X-ray observations}

\author[Lira, Lawrence \& Johnson]
{P.~Lira,$^1$ A.~Lawrence,$^2$ R.A.~Johnson,$^3$\\
$^1$ Department of Physics \& Astronomy, University of Leicester, 
Leicester LE1 7RH, UK\\
$^2$ Institute for Astronomy, University of Edinburgh, Royal Observatory, 
Blackford Hill, Edinburgh EH9 3HJ, Scotland\\
$^2$ Institute of Astronomy, Madingley Road, Cambridge CB3 0HA, UK\\}

\begin{document}

\maketitle

\begin{abstract}

We discuss ROSAT HRI X-ray observations of 33 very nearby galaxies,
sensitive to X-ray sources down to a luminosity of approximately
$10^{38}$ erg s$^{-1}$. The galaxies are selected from a complete,
volume limited sample of 46 galaxies with $d < 7$ Mpc for which we
have extensive multi-wavelength data. For an almost complete
sub-sample with $M_{B} < -14$ (29/31 objects) we have HRI
images. Contour maps and source lists are presented within the central
region of each galaxy, together with nuclear upper limits where no
nuclear source was detected. Nuclear X-ray sources are found to be
very common, occurring in $\sim 35\%$ of the sample.  Nuclear X-ray
luminosity is statistically connected to host galaxy luminosity -
there is not a tight correlation, but the probability of a nuclear
source being detected increases strongly with galaxy luminosity and
the distribution of nuclear luminosities seems to show an upper
envelope that is roughly proportional to galaxy luminosity. While
these sources do seem to be a genuinely nuclear phenomenon rather than
nuclear examples of the general X-ray source population, it is far
from obvious that they are miniature Seyfert nuclei. The more luminous
nuclei are very often spatially extended, and HII region nuclei are
detected just as often as LINERs. Finally, we also note the presence
of fairly common super-luminous X-ray sources in the off-nuclear
population -- out of 29 galaxies we find 9 sources with a luminosity
larger than 10$^{39}$ erg s$^{-1}$. These show no particular
preference for more luminous galaxies. One is already known to be a
multiple SNR system, but most have no obvious optical counterpart and
their nature remains a mystery.

\end{abstract}


\section{Introduction}

Evidence has been growing that both (i) weak nuclear activity, and
(ii) the presence of quiescent black holes, are common features of
normal galaxies. The best evidence to date for (i) comes from the very
thorough high S/N spectroscopic survey of 486 nearby galaxies by
\scite{hfs}. They found that 86\% of galaxies have emission lines;
roughly a third have Seyfert 2 or LINER spectra, indicating but not
proving some kind of weak AGN; and 10\% have weak broad emission
lines, so are almost certainly AGN \cite{hfs3}. These latter objects
are, however, three orders of magnitude less luminous than classic
Seyfert galaxies and six orders of magnitude less luminous than the
most luminous quasars.  They have become known as `dwarf AGN'. The
evidence for (ii) comes from dynamical studies. The review by
\scite{kormendy+richston} found evidence for `Massive Dark Objects'
(MDOs) in 20\% of galaxies, but this can be considered a lower limit
because of the difficulty of rigorously demonstrating the requirement
for a dark mass. With a more relaxed analysis, \scite{magorrian+etal}
have claimed that there is strong evidence for an MDO in essentially
all galaxies.

Most interestingly, \scite{kormendy+richston} and
\scite{magorrian+etal} also claim a correlation between the MDO mass
and the stellar bulge mass, with $M_{MDO}/M_{bulge} \sim 0.003 -
0.006$. Likewise, various authors have shown evidence that AGN
activity correlates with galaxy size
\cite{gehren+etal,hutchings+etal,dibai+zasov,sadler,huchra+burg,mcleod3}. A
weakness with all these studies is that only fairly luminous galaxies
are considered.  As well as investigating whether feeble AGN are
present in normal galaxies, it is very important to know whether AGN
occur at all in feeble galaxies. Is there a minimum galaxy size below
which quasar-like nuclei are simply not present? There are two known
examples of AGN in relatively small galaxies: a Seyfert 1 nucleus in
NGC\,4395 \cite{filippenko+sargent,lira} and a Seyfert 2 nucleus in
G1200-2038 \cite{kunth+etal}. Both these galaxies have $M_{B} \sim
-18$, an order of magnitude less luminous than typical $L^{\ast}$
galaxies and two orders of magnitude less luminous than the giant
ellipticals which dominate the \scite{magorrian+etal} sample. Are they
rare freaks, or the tip of the iceberg? Furthermore, in the local
field there are are plenty of spiral galaxies an order of magnitude
less luminous still than NGC\,4395 and below that we have true dwarf
galaxies, dwarf irregulars, dwarf ellipticals, and dwarf spheroidals,
which continue all the way down to $M_{B} \sim -7$.  Where does AGN
activity stop?

To address the issues of (a) how common weak AGN are, and (b) how
activity is connected to host galaxy size and type, we have undertaken
a multi-wavelength study of a complete volume-limited sample of 46
very nearby AGN, with (broad-band and emission line) optical, IR,
X-ray and radio imaging, and long slit spectroscopy. In terms of
survey size and spectroscopic quality it would be hard to improve on
the \scite{hfs} survey. Instead, our study has several key
features. (i) The multiwavelength approach gives the greatest chance
of finding a weak AGN, being sensitive to a variety of signatures, and
gives us diagnostic information on the nature of the objects seen.
(ii) All our galaxies are closer than 7 Mpc, maximizing spatial
resolution, and minimizing the detectable luminosity. (iii) The
volume-limited nature (as opposed to magnitude-limited) of the sample
means that we have a fairly even distribution of galaxy sizes and
types, going down to extremely small galaxies. (iv) In many cases we
found that the location of the true nucleus is not entirely
obvious. Multi-wavelength imaging gave us a `shooting list' of
multiple targets for spectroscopy to ensure we did not miss the true
nucleus.

This paper (Paper I) focuses on the analysis of the ROSAT HRI X-ray
data for the galaxies in our sample and presents some preliminary
scientific results.  Section 2 describes the main characteristics of
the sample. In section 3 the acquisition and analysis of the X-ray
data are described. The atlas of X-ray images is found in section
4. In Section 5 we describe detailed results for individual
galaxies. In section 6 we summarize and discuss some of the most
interesting results and examine the correlation of X-ray luminosity
and host galaxy luminosity.

In subsequent papers we will present optical-IR imaging and radio maps
(Paper II), long-slit spectroscopy (Paper III), and a final analysis
of results (Paper IV).

\section{The volume-limited sample}

\begin{table*}
\caption{The volume-limited sample of galaxies. The available data 
(as of September 1999) are indicated with a \y. Data that should
become available in the near future are indicated with a \w. $a$
LEO\,B was observed with the Einstein HRI only.}
\label{tab:allobs}
\centering
\begin{tabular}[t]{lc@{\hspace{0.3cm}}r@{\hspace{0.3cm}}ccccccc} \hline
 &&&&& \multicolumn{4}{c}{Broad Band Observations} & Nuclear\\
 Galaxy & Position (2000) & $-M_{B}$ & Morph & D (Mpc) & Radio & Near-IR & Optical & X-ray & Spect.\\ [10pt]	 
 NGC\,147 		&00 30  +48 14	 	&13.71 		&dE5        &0.65 	&\y &\y &\y &\y &\y \\	
 AND\,III    		&00 33  +36 14    	&10.17 		&dE         &0.65 	&\y &   &\y &   &   \\	
 NGC\,185    		&00 36  +48 04 		&13.94 		&dE3        &0.65 	&\y &\y &\y &\y &\y \\	
 NGC\,205--M\,110   	&00 38  +41 25 		&15.24		&S0/E5      &0.65 	&\y &\y &\y &\y &\y \\	
 NGC\,221--M\,32   	&00 40  +40 36  	&15.06 		&E2         &0.65 	&\y &\y &\y &\y &\y \\	
 NGC\,224--M\,31    	&00 40  +40 60  	&19.69 		&Sb         &0.65 	&\y &\y &\y &\y &\y \\	
 AND\,I      		&00 43  +37 44  	&10.17 		&dE3        &0.65 	&\y &\y &\y &   &   \\	
 NGC\,247     		&00 45  -21 02  	&18.32 		&Sc(s)      &3.69 	&\y &   &\y &\y &\y \\	
 NGC\,253      		&00 45  -25 34  	&20.26 		&Sc(s)      &4.77 	&\y &\y &   &\y &   \\ 	
 SCULPTOR		&00 58  -33 58  	&10.38		&dE3        & -- 	&\y &   &   &   &   \\	
 NGC\,404    		&01 07  +35 27 		&17.07  	&S0         &4.34 	&\y &\y &\y &\y &\y \\	
 AND\,II        	&01 14  +33 09  	&10.17 		&dE         &0.65 	&\y &\y &\y &   &   \\	
 NGC\,598--M\,33    	&01 31  +30 24  	&18.43 		&Sc(s)      &0.87 	&\y &\y &   &\y &\y \\	
 MAFFEI\,I   		&02 33  +59 26  	&13.87 		&E3         &5.43 	&\y &\y &\y &   &\y \\	
 FORNAX    		&02 38  -34 44  	&12.64 		&dE0        &0.22 	&\y &   &   &   &   \\	
 MAFFEI\,II    		&02 38  +59 23 		& 8.08  	&Sbc        &5.21 	&\y &\y &\y &   &\y \\	
 IC\,342    		&03 42  +67 56  	&19.50 		&S(B)cd(rs) &6.08 	&\y &\y &\y &\y &\y \\	
 NGC\,1560 		&04 27  +71 46		&15.95		&Sd(s)      &4.34 	&\y &   &   &\y &\y \\	
 NGC\,2366   		&07 24  +69 19 		&17.53  	&SBm        &6.29 	&\y &\y &\y &\y &\y \\	
 NGC\,2403   		&07 32  +65 43  	&20.25 		&Sc(s)      &6.73 	&\y &\y &\y &\y &\y \\	
 NGC\,2976 		&09 43  +68 09 		&16.98  	&Sd         &3.69 	&\y &\y &\y &\y &\y \\	
 A\,0951+68  		&09 51  +68 50  	&11.21 		&dE         &1.52 	&\y &\y &\y &\y &\y \\	
 NGC\,3031--M\,81   	&09 52  +69 18  	&19.22 		&Sb(r)      &2.60 	&\y &\y &\y &\y &   \\	
 LEO\,I     		&10 06  +12 33 		&10.87 		&dE3        &0.22 	&\w &\y &   &   &   \\	
 LEO\,B$^{a}$     	&11 11  +22 26 		& 8.98 		&dE0        &0.22 	&\w &\y &   &\y &\y \\	
 UGC\,6456    		&11 25  +79 16  	&12.19 		&peculiar   &2.39 	&\y &   &\y &\y &\y \\	
 NGC\,3738  		&11 33  +54 48 		&16.78  	&Sd         &6.08 	&\w &\y &\y &\y &\y \\	
 NGC\,4136 		&12 07  +30 12 		&17.63  	&Sc(r)      &6.94 	&\y &\y &\y &\y &   \\	
 NGC\,4144		&12 07  +46 44 		&16.83  	&Scd        &5.86 	&\w &\y &\y &\y &\y \\	
 NGC\,4150   		&12 08  +30 41 		&15.30  	&S0/a       &3.47 	&\y &\y &\y &\y &\y \\	
 NGC\,4236   		&12 14  +69 45 		&17.50  	&SBd        &3.26 	&\w &\y &\y &\y &\y \\	
 NGC\,4244   		&12 15  +38 05 		&17.59  	&Scd        &4.34 	&\y &\y &\y &\y &   \\	
 UGC\,7321      	&12 15  +22 49  	&14.82 		&Scd        &5.43 	&\w &   &   &   &\y \\	
 NGC\,4395   		&12 23  +33 50 		&17.89  	&Sd         &5.21 	&\y &\y &   &\y &\y \\	
 NGC\,4605   		&12 38  +61 53 		&17.80  	&Sc(s)      &5.64 	&\w &\y &\y &\y &\y \\	
 NGC\,4736--M\,94   	&12 49  +41 24  	&20.00 		&RSab(s)    &6.08 	&\y &\y &\y &\y &\y \\	
 NGC\,4826   		&12 54  +21 57  	&19.39 		&Sab(s)     &5.64 	&\w &\y &\y &\y &\y \\	
 NGC\,5204      	&13 28  +58 41 		&17.39  	&Sd         &6.73 	&\w &\y &\y &\y &\y \\	
 NGC\,5238   		&13 33  +51 52  	&15.79 		&S(B)dm     &6.73 	&\w &\y &\y &\y &\y \\	
 NGC\,5236--M\,83   	&13 34  -29 37 		&20.16 		&SBc(s)     &5.43 	&\y &\y &\y &\y &   \\	
 NGC\,5457--M\,101   	&14 01  +54 36  	&21.22 		&Sc(s)      &7.60 	&\y &\y &\y &\y &\y \\	
 URSA\,MINOR    	&15 08  +67 23 		& 7.40 		&dE4        & -- 	&   &   &\y &   &   \\	
 DRACO  		&17 19  +57 58 		& 7.00 		&dE0        & -- 	&   &\y &\y &   &   \\
 NGC\,6503   		&17 50  +70 10 		&18.28  	&Sc(s)      &6.94 	&\y &\y &\y &\y &\y \\	
 NGC\,6946   		&20 34  +59 59  	&19.66 		&Sc(s)      &7.38 	&\y &\y &\y &\y &\y \\	
 NGC\,7793     		&23 55  -32 52  	&18.43 		&Sd(s)      &4.12 	&\y &\y &   &   &   \\	
\hline
\end{tabular}
\end{table*}

To study the occurrence of nuclear activity in galaxies to the lowest
possible level, a volume-limited sample of galaxies was defined.
The sample contains the nearest examples of various morphological
types of galaxies, and a representative range of intrinsic
luminosities (Johnson 1997; see also Paper II).

The starting point was the Kraan-Korteweg \& Tamman Catalogue
\cite{kraan-korteweg+tamman,kraan-korteweg} with all galaxies with V
$<$ 500 km/s giving a distance-complete sample within $d=0.35 \times
d_{\rm Virgo}$ (after applying a virgocentric flow model with an
infall velocity of the Local Group equal to 220 km s$^{-1}$). With
$d_{\rm Virgo} = 21.5$ Mpc, the sample is complete for nearby galaxies
within 7.6 Mpc.  The catalogue is not complete for intrinsically faint
galaxies without velocity information, but it includes all galaxies
down to $M_{B} \sim -13$.  Comparing the distribution of absolute
magnitudes of the galaxies in the KKT with the galaxies in the Revised
Shapley Ames (magnitude-limited) catalogue \cite{sandage+tamman} it is
found that the RSA sample of galaxies presents a strong peak around
$M_{B}=-21$ and has less than 10\% of the objects below $M_{B}=-18$,
while the KKT sample shows a broad distribution with a maximum around
$M_{B}=-16$ and a significant number of galaxies down to magnitude
$M_{B}=-13$.

The observations were restricted to objects with $\delta >
-35^{\circ}$. Because the aim was to observe objects with well-formed
nuclei, galaxies classified as Sdm, Sm and Irr were removed from the
sample.  This gave a basic sample of 46 nearby galaxies (26 spirals, 5
E/S0, 12 dwarf ellipticals and 3 blue compact dwarfs, of which two are
classified as Sm spirals and one is a peculiar object). The assumed
distances for the galaxies are based on the observed recession
velocities and the assumed Virgo flow model. 

Table \ref{tab:allobs} shows the resulting sample on which our study
is based, the properties of the galaxies concerned, and a summary of
the observations available to date. Note that only 34/46 galaxies have
available X-ray data. However, we can define a very useful sub-sample
of all those galaxies with $M_{B} < - 14$ (these are all disk systems,
except for the elliptical galaxy NGC\,221). Of these, 29/31 have X-ray
data -- only UGC\,7321 and NGC\,7793 are missing. In addition this
paper describes observations of five further dwarf galaxies: NGC\,147,
NGC\,185, UGC\,6456, Leo\,B, and A\,0951+68. Only one of these has a
detected X-ray source. The analysis we present in section 6 is based
on the almost-complete subsample with $M_{B} < -14$.

\section{High resolution X-ray imaging}

\subsection{Observations}

The data reported here come from pointed observations carried out with
the ROSAT HRI. Table \ref{tab:hriobs} shows those galaxies in the
volume-limited sample with available data. Observations for 17
galaxies were awarded to this project through Announcement of
Opportunity (AO) calls.  For 13 objects data were retrieved from the
public archives using {\sc arnie} (a World Wide Web interface to the
databases and catalogues supported by the Leicester Database and
Archive Service, {\sc ledas}).  Galaxies are identified in table
\ref{tab:hriobs} with an `N' (new data allocated to this project) or a
`P' (public data retrieved from the archive). Since NGC\,4826 has both
new and archive observations a total of 29 objects are covered.  The
ROSAT HRI data have been reduced and analyzed as described below.

\begin{table*}
\caption{ROSAT HRI observations. The origin of the data is marked
as `N' for new data allocated to this project, and `P' for public 
data retrieved from the archives.}
\label{tab:hriobs}
\centering
\begin{tabular}{lccrclccrc} \hline 
Galaxy & ROR Number & Observing Date & Livetime & Origin & Galaxy & ROR Number & Observing Date & Livetime & Origin\\ [10pt]
NGC\,147  & 400744 & 19/01/95 &	14616.6 & P & NGC\,4236 & 600763 & 01/12/94 & 7543.2  & N \\
NGC\,185  & 400743 & 19/01/95 &	20881.3 & P &           & 600763 & 10/04/95 & 3390.6  & N \\
NGC\,205  & 600816 & 04/08/96 &	27842.5 & P & NGC\,4244 & 702724 & 20/06/96 & 8630.9  & N \\
NGC\,221  & 600600 & 19/07/94 &	12533.5 & P & NGC\,4395 & 702725 & 23/06/96 & 11252.5 & N \\
NGC\,247  & 600622 & 29/06/94 & 33701.1 & P & NGC\,4605 & 702729 & 16/06/96 & 2158.4  & N \\
          & 600622 & 12/06/95 & 17681.8 & P & NGC\,4736 & 600678 & 07/12/94 & 111870.5& P \\
NGC\,404  & 703894 & 04/01/97 &	23562.6 & N &           & 600769 & 25/12/94 & 27033.8 & P \\
IC\,342   & 600022 & 13/02/91 &	18990.4 & P & NGC\,4826 & 600715 & 09/07/95 & 10042.8 & P \\
NGC\,1560 & 702727 & 08/03/96 &	17287.5 & N &           & 703900 & 12/06/97 & 9248.9  & N \\
NGC\,2366 & 702732 & 31/03/96 &	31465.5 & N &           & 703900 & 08/01/98 & 5342.2  & N \\
NGC\,2403 & 600767 & 18/09/95 &	26244.7 & P & NGC\,5204 & 702723 & 31/12/95 & 14683.0 & N \\
NGC\,2976 & 600759 & 04/10/94 & 26238.1 & P &           & 702723 & 02/05/96 & 13554.1 & N \\
          & 600759 & 12/04/95 & 23244.3 & P & NGC\,5236 & 600024 & 20/01/93 & 23316.5 & P \\
A\,0951+68& 703895 & 29/03/97 &	14240.2 & N &           & 600024 & 30/07/94 & 23920.5 & P \\
UGC\,6456 & 703896 & 01/04/97 & 13027.7 & N & NGC\,5238 & 702733 & 13/06/96 & 23163.1 & N \\
NGC\,3738 & 703897 & 01/11/97 & 13711.2 & N & NGC\,5457 & 600092 & 09/01/92 & 18452.8 & P \\
NGC\,4136 & 702734 & 25/06/96 & 1951.0  & N &           & 600383 & 10/12/92 & 32361.1 & P \\
NGC\,4144 & 703898 & 11/11/97 & 11083.4 & N & NGC\,6503 & 600618 & 08/03/94 & 14640.6 & N \\
NGC\,4150 & 600762 & 25/12/94 & 3730.1  & N & NGC\,6946 & 600501 & 14/05/94 & 59885.3 & P \\
          & 600762 & 23/05/95 & 10353.8 & N &           & 600718 & 13/08/95 & 21514.3 & P \\
\hline
\end{tabular}
\end{table*}

For several galaxies more than one exposure was available.  This could
be because several observations were requested, or because a single
request was scheduled as several observations months apart. In table
\ref{tab:hriobs} observations of the same galaxy associated with
different requests are distinguished by their ROR (ROSAT Observation
Request) numbers. If a single request was scheduled in more than one
observation only one ROR is given but the different observing dates
distinguish between the individual observations.

\subsection{Data analysis}

The aim of the analysis is to extract fluxes and upper limits for all
the sources within a $\sim 6\arcmin \times 6 \arcmin$ region centered
on the position of the galaxy. In this way the X-ray field matches the
size of the optical frames obtained using the Jacobus Kapteyn
Telescope (JKT) (Johnson 1997; Paper II). For each galaxy an
isointensity contour map was overlaid onto the optical image. No
attempt has been made to systematically study the temporal behavior of
the sources, although some interesting individual cases will be
mentioned later (section 5).

The reduction and analysis of the ROSAT HRI data was done using {\sc
pros}, an X-ray analysis software system designed to run under the
Image Reduction and Analysis Facility ({\sc iraf}).

For each galaxy with more than one exposure, the different images were
coadded, merging their respective livetimes.  Using the sources in the
field, the images were inspected and corrected for systematic shifts
in the positions before co-adding. 

\subsubsection{Source identification}

For the identification of all point sources in the images we used the
ROSAT Standard Analysis Software System (SASS) source list as a
starting point. The following steps were followed to identify all
source candidates in the X-ray images:

\begin{enumerate}

\item the images were binned into 2\arcsec\ pixels and smoothed 
using gaussians with $\sigma = 2\arcsec, 4\arcsec$ and $8\arcsec$; the
smoothed images were inspected visually to evaluate the existence of
sources not reported by SASS and to exclude obvious spurious
sources;

\item whenever more than one exposure was available for a target the 
different files were compared so that variable sources that might
appear weak in the coadded image could be recognized in the individual
frames where they might have been more luminous;

\item optical images with X-ray isocontours were generated in order to 
look for weak sources with optical counterparts.

\end{enumerate}

For each source candidate a background subtracted count number was
obtained from the coadded imaged. An aperture of 10\arcsec\ was
adopted for all point sources which should encircle $\sim$ 99\% of the
photons at 0.2 keV and $\sim$ 86\% of the photons at 1.7 keV for
nearly on-axis sources \cite{david+etal}. The pixel coordinates were
obtained from the SASS report or by using a separate centroid
algorithm. To estimate the background one or two large circular
regions free of evident X-ray sources and away from the galaxy were
used. These regions normally lay outside the $\sim 6\arcmin \times 6
\arcmin$ central image.

A final list of sources for each galaxy was created with all sources
that comply with at least one of the following criteria:

\begin{enumerate}

\item have a signal-to-noise ratio above 2.5 in the coadded image,

\item have a signal-to-noise ratio above 2.5 in at least one of the 
individual images,

\item or have a low signal-to-noise ratio (between 1.5 and 2.5) and
an optical counterpart.

\end{enumerate}

For sources with an extended component an additional, larger aperture
was used to estimate its contribution. The size of the aperture was
determined from the radial profile of the source.  For galaxies
without nuclear X-ray source detections, upper limits were established
using a 10\arcsec\ aperture located at the best estimate of the
nuclear position from our optical images. They were computed as
$2\sigma$ limits (ie., 0.9772 CL) assuming Bayesian statistics
\cite{kraft}. To find the count rates the total livetime of the
coadded images was used.

\begin{table*}
\caption{Factors to convert fluxes obtained assuming Bremsstrahlung 
emission with $kT = 5.0$ to Bremsstrahlung emission with $kT = 0.1$ or
1.0 keV, or a power law model with index $\alpha = 1.5$ or 1.0: New
flux = flux ($kT = 5.0) \times$ Conversion Factor.}
\label{tab:nh_conv}
\centering
\begin{tabular}{l@{\hspace{0.6cm}}cr@{\hspace{0.6cm}}ccc@{\hspace{0.6cm}}c@{\hspace{0.6cm}}} \hline 
Galaxy & Size & $\log N_{H}$ & \multicolumn{2}{c}{Bremsstrahlung} & \multicolumn{2}{c}{Power Law} \\
       &(deg$^{2}$)&(cm$^{-2}$)  & $kT = 0.1$ keV & $kT = 1.0$ keV & $\alpha = 1.5$ & $\alpha = 1.0$ \\ [10pt]
NGC\,147  & 0.091 & 21.02 & 24.71 & 1.21 & 2.26 & 1.42 \\
NGC\,185  & 0.098 & 21.04 & 25.02 & 1.22 & 2.27 & 1.42 \\
NGC\,205  & 0.148 & 20.82 & 10.87 & 1.17 & 2.07 & 1.37 \\
NGC\,221  & 0.038 & 20.81 & 10.58 & 1.17 & 2.07 & 1.37 \\
NGC\,247  & 0.129 & 20.17 &  1.96 & 1.01 & 1.36 & 1.14 \\
NGC\,404  & 0.016 & 20.70 &  8.28 & 1.15 & 1.98 & 1.34 \\
IC\,342   & 0.270 & 21.48 & 43.83 & 1.30 & 2.48 & 1.48 \\
NGC\,1560 & 0.017 & 21.06 & 25.34 & 1.22 & 2.27 & 1.43 \\
NGC\,2366 & 0.023 & 20.59 &  4.62 & 1.11 & 1.78 & 1.29 \\
NGC\,2403 & 0.170 & 20.62 &  5.23 & 1.12 & 1.83 & 1.30 \\
NGC\,2976 & 0.011 & 20.65 &  6.05 & 1.13 & 1.89 & 1.32 \\
A\,0951+68&  --   & 20.63 &  5.48 & 1.12 & 1.85 & 1.31 \\
UGC\,6456 & 0.001 & 20.56 &  4.15 & 1.10 & 1.74 & 1.27 \\
NGC\,3738 & 0.004 & 20.01 &  1.67 & 0.99 & 1.27 & 1.10 \\
NGC\,4136 & 0.014 & 20.20 &  2.03 & 1.02 & 1.38 & 1.15 \\
NGC\,4144 & 0.008 & 20.16 &  1.94 & 1.01 & 1.36 & 1.14 \\
NGC\,4150 & 0.004 & 20.19 &  2.01 & 1.02 & 1.38 & 1.15 \\
NGC\,4236 & 0.112 & 20.26 &  2.20 & 1.03 & 1.43 & 1.16 \\
NGC\,4244 & 0.036 & 20.27 &  2.23 & 1.03 & 1.43 & 1.17 \\
NGC\,4395 & 0.123 & 20.12 &  1.85 & 1.00 & 1.33 & 1.13 \\
NGC\,4605 & 0.011 & 20.26 &  2.20 & 1.03 & 1.43 & 1.16 \\
NGC\,4736 & 0.087 & 20.15 &  1.92 & 1.01 & 1.35 & 1.13 \\
NGC\,4826 & 0.044 & 20.42 &  2.88 & 1.06 & 1.57 & 1.22 \\
NGC\,5204 & 0.013 & 20.18 &  1.98 & 1.01 & 1.37 & 1.14 \\
NGC\,5236 & 0.100 & 20.63 &  5.48 & 1.12 & 1.85 & 1.31 \\
NGC\,5238 & 0.100 & 20.04 &  1.71 & 0.99 & 1.29 & 1.11 \\
NGC\,5457 & 0.618 & 20.07 &  1.76 & 1.00 & 1.30 & 1.11 \\
NGC\,6503 & 0.013 & 20.61 &  5.01 & 1.12 & 1.82 & 1.30 \\
NGC\,6946 & 0.094 & 21.31 & 31.90 & 1.26 & 2.37 & 1.45 \\
\hline
\end{tabular}
\end{table*}

\subsubsection{Astrometry checks}

In order to check the absolute astrometry of the HRI observations we
used bright X-ray sources in the field of view of each image and
looked for optical counterparts that would signal the presence of
large shifts in the astrometric solutions. However, in many cases the
only bright sources available were found at large off-axis
angles. Given the dependancy of the PSF with distance from the centre
of the HRI optical axis \cite{david+etal}, it is important to
establish if the position of these sources could be reliably
recovered.

To establish the accuracy with which the position of X-ray sources
could be recovered we examined two HRI observations for which a large
number of optical couterparts for the X-ray sources has been found: one
of the Lockman Hole observations (ROR 701867) and one of the ROSAT
Deep Surveys centered in the direction $\alpha = 13^{h} 34^{m} 36^{s}$
and $\delta = 37\degr 54\arcmin 36\arcsec$ (ROR 900717). The optical
identifications of the X-ray sources were obtained from
\scite{schmidt+etal} and \scite{mchardy+etal}.

We used a total of 12 sources from the Lockman Hole and the ROSAT Deep
Survey image, with off-axis angles ranging from 4 to 14.5 arcmin, and count
numbers ranging from $\la$ hundred to a few thounsands. It was found
that the position of the X-ray sources could be recovered with an
accuracy better than 5 arcsec (average of 2.8 arcsec) regardless of
the count numbers and off-axis angle of the source. Therefore, by using
optical identifications (IDs) of the X-ray sources found in the field
of view of our HRI observations, it should be possible to detect
shifts in the astrometric solutions larger than 5 arcsec.

We checked the astrometry of our HRI observations for all those images
that showed X-ray sources that could be associated with the
galaxies. To look for optical counterparts large DSS images were used
and inspected carefully to identify as many candidate IDs as
possible. The celestial positions for the IDs were obtained usign a
centroid algorithm and compared with the positions of the X-ray
sources. The results can be grouped into two categories as follows:
fields for which at least two secure optical IDs were found (NGC\,247,
NGC\,2976, NGC\,5457, NGC\,6503, NGC\,6946, UGC\,6456); fields for
which only one secure ID was found (IC\,342, NGC\,221, NGC\,404,
NGC\,2403, NGC\,4136, NGC\,4150, NGC\,4236, NGC\,4395, NGC\,4736,
NGC\,4826, NGC\,5204, NGC\,5236).

The analysis showed that most astrometric solutions were accurate to
within 5 arcsec. Exceptions were NGC\,2976, for which the observation
obtained on the 12/04/1995 presented a shift of $\sim 6$ arcsec,
NGC\,4236 for which both observations presented a significant shift
(notice however, that this result is based on only one optical ID),
and NGC\,6946 -- ROR 600718, which was corrected by a shift $\sim 7$
arcsec.

\subsubsection{X-ray fluxes and luminosities}

The conversion of the HRI count rates to fluxes was done assuming a
Bremsstrahlung spectrum with $kT = 5$ keV, an energy range 0.1--2.4
keV, and a Galactic line of sight absorption derived from the 21 cm
line of atomic hydrogen \cite{stark+etal}, listed in table
\ref{tab:nh_conv}. This choice of parameters is justified by typical
galactic soft X-ray spectral properties
\cite{kim+fabbiano+trinchieri}.  Fluxes were converted to luminosities
assuming the distances in table \ref{tab:allobs}. Fluxes and
luminosities inferred from a Bremsstrahlung spectrum with $kT = 1.0$
or 0.1 keV, or a power law model with index $\alpha = 1.0$ or 1.5, can
be calculated using the conversion factors in table \ref{tab:nh_conv}
for each galaxy.

\subsection{Contour maps}

Isointensity contour maps were produced for all the coadded
images. All the files were binned into $2\arcsec \times 2\arcsec$
pixels and then smoothed using a gaussian with $\sigma = 4\arcsec$.
The binning and gaussian sizes were chosen so that the representation
of the sources was consistent with the statistical asssesment of their
strength. Contours were drawn at $2.5^{n}$ times the standard
deviation in the smoothed background, where $n = 1, 2, 3$, and so
on. With this selection of contour intensities bright sources do not
present contour crowding. For each map, the background fluctuation was
calculated for the same region used earlier (when measuring source
fluxes in the raw data).

\section{The atlas}

\begin{table*}
\caption{ROSAT HRI fluxes and luminosities. Continued over.}
\label{tab:fluxes}
\begin{small}
\centering
\begin{tabular}{c@{\hspace{0.6cm}}c@{\hspace{0.6cm}}c@{\hspace{0.6cm}}ccl} \hline
Galaxy 		& Source & S/N & Corrected Flux & Luminosity & Comments \\ 
       		&        &     &(ergs s$^{-1}$ cm$^{-2}$)& (ergs s$^{-1}$) &\\ [10pt]
NGC\,205 	& X1 & 3.82  & $7.93 \ti 10^{-14}$  &  $4.01 \ti 10^{36}$ & Probably not associated with NGC\,205\\[6pt]
NGC\,221 	& X1 & 10.60 & $6.37 \ti 10^{-13}$ &  $3.22 \ti 10^{37}$ &\\[6pt]
NGC\,247 	& X1 & 22.39 & $7.37 \ti 10^{-13}$ &  $1.20 \ti 10^{39}$ &\\
         	& X2 &  5.99 & $9.15 \ti 10^{-14}$ &  $1.49 \ti 10^{38}$ &\\[6pt]
NGC\,404 	& X1 &  3.34 & $7.48 \ti 10^{-14}$ &  $1.69 \ti 10^{38}$ & Nuclear\\[6pt]
IC\,342$^{a}$ 	& X1 &  2.31 & $9.58 \ti 10^{-14}$ &  $4.24 \ti 10^{38}$ & Low S/N (see text); source 6 in BCT\\
		& X2 &  9.27 & $6.07 \ti 10^{-13}$ &  $2.68 \ti 10^{39}$ & Nuclear; marginally extended; source 8 in BCT\\
		& X3 &  6.48 & $3.44 \ti 10^{-13}$ &  $1.52 \ti 10^{39}$ & Possible optical counterpart; source 9 in BCT\\[6pt]
NGC\,2403	& X1 & 15.19 & $5.56 \ti 10^{-13}$ &  $3.01 \ti 10^{39}$ &\\
		& X2 &  2.92 & $5.67 \ti 10^{-14}$ &  $3.07 \ti 10^{38}$ &\\
		& X3 &  7.88 & $1.89 \ti 10^{-13}$ &  $1.02 \ti 10^{39}$ & Possible very faint optical counterpart\\
		& X4 &  2.08 & $4.29 \ti 10^{-14}$ &  $2.33 \ti 10^{38}$ & Low S/N; giant HII region \cite{drissen+roy}\\[6pt]
NGC\,2976	& X1 &  8.75 & $1.32 \ti 10^{-13}$ &  $2.15 \ti 10^{38}$ &\\[6pt]
A\,0951+68	& X1 &  4.79 & $1.06 \ti 10^{-13}$ &  $4.00 \ti 10^{37}$ & Probably not associated with A\,0951+68\\[6pt]
UGC\,6456	& X1 & 13.85 & $8.82 \ti 10^{-13}$ &  $6.03 \ti 10^{38}$ &\\[6pt]
NGC\,3738	& X1 &  4.74 & $1.18 \ti 10^{-13}$ &  $5.22 \ti 10^{38}$ & Probably not associated with NGC\,3738\\
		& X2 &  2.46 & $5.25 \ti 10^{-14}$ &  $2.32 \ti 10^{38}$ & Probably not associated with NGC\,3738\\[6pt]
NGC\,4136	& X1 &  2.15 & $2.19 \ti 10^{-13}$ &  $1.26 \ti 10^{39}$ & Diffuse blue optical counterpart\\[6pt]
NGC\,4144	& X1 &  3.09 & $8.77 \ti 10^{-14}$ &  $3.60 \ti 10^{38}$ &\\[6pt]
NGC\,4150	& X1 & 15.84 & $1.02 \ti 10^{-12}$ &  $1.88 \ti 10^{45}$ & Background quasar ($z=0.52$); see section 5;\\
		&&&&&							flux obtained assuming a power law spectrum\\ 
		&&&&&							with $\alpha=1.0$\\ [6pt]
NGC\,4236	& X1 &  2.61 & $7.24 \ti 10^{-14}$ &  $9.21 \ti 10^{37}$ &\\[6pt]
NGC\,4395	& X1 &  1.67 & $3.73 \ti 10^{-14}$ &  $1.21 \ti 10^{38}$ & Nuclear; low S/N; flux obtained assuming a\\
		&&&&&						     power law spectrum with $\alpha=1.0$; see chapter 4\\[6pt]
		& X2 & 11.85 & $6.46 \ti 10^{-13}$ &  $2.10 \ti 10^{39}$ &\\
NGC\,4736	& X1 &  5.78 & $3.40 \ti 10^{-14}$ &  $1.51 \ti 10^{38}$ &\\
		& X2 & 18.78 & $1.46 \ti 10^{-13}$ &  $6.46 \ti 10^{38}$ &\\
		& X3 & 65.32 & $1.38 \ti 10^{-12}$ &  $6.11 \ti 10^{39}$ & Nuclear\\
		& X3 & 84.49 & $4.70 \ti 10^{-12}$ &  $2.08 \ti 10^{40}$ & Total nuclear emission (r = 100\arcsec)\\
NGC\,4826	& X1 &  9.58 & $2.41 \ti 10^{-13}$ &  $9.19 \ti 10^{38}$ & Nuclear\\
		& X1 & 13.03 & $7.73 \ti 10^{-13}$ &  $2.94 \ti 10^{39}$ & Total nuclear emission (r = 40\arcsec) \\
		& X2 &  2.27 & $3.98 \ti 10^{-14}$ &  $1.52 \ti 10^{38}$ & Probably not associated with NGC\,4826\\[6pt]
NGC\,5204	& X1 & 25.23 & $1.08 \ti 10^{-12}$ &  $5.86 \ti 10^{39}$ & Faint optical counterpart\\[6pt]
NGC\,5236	& X1 &  4.46 & $5.83 \ti 10^{-14}$ &  $2.06 \ti 10^{38}$ & Variable\\ 			
		& X2 &  6.07 & $8.23 \ti 10^{-14}$ &  $2.90 \ti 10^{38}$ &\\ 				
		& X3 & 26.69 & $8.75 \ti 10^{-13}$ &  $3.09 \ti 10^{39}$ & Nuclear\\ 			
		& X3 & 38.48 & $3.01 \ti 10^{-12}$ &  $1.06 \ti 10^{40}$ & Total nuclear emission (r = 100\arcsec)\\
		& X4 &  5.09 & $6.69 \ti 10^{-14}$ &  $2.36 \ti 10^{38}$ &\\				
		& X5 &  3.78 & $4.95 \ti 10^{-14}$ &  $1.75 \ti 10^{38}$ &\\
		& X6 &  8.60 & $1.31 \ti 10^{-13}$ &  $4.61 \ti 10^{38}$ &\\				
		& X7 &  5.01 & $8.45 \ti 10^{-14}$ &  $2.98 \ti 10^{38}$ & Variable\\			
		& X8 &  4.13 & $5.38 \ti 10^{-14}$ &  $1.90 \ti 10^{38}$ & Variable\\[6pt] 		
NGC\,5457	& X1 &  4.13 & $3.88 \ti 10^{-14}$ &  $2.68 \ti 10^{38}$ & Nuclear\\
		& X2 &  3.52 & $3.32 \ti 10^{-14}$ &  $2.30 \ti 10^{38}$ &\\
		& X3 &  4.80 & $4.56 \ti 10^{-14}$ &  $3.15 \ti 10^{38}$ &\\
		& X4 &  3.31 & $3.16 \ti 10^{-14}$ &  $2.19 \ti 10^{38}$ &\\
\hline
\end{tabular}
\end{small}
\end{table*}

\begin{table*}
\contcaption{ROSAT HRI fluxes and luminosities. $a$: Re-analysis of the observations 
reported by \protect\scite{bregman}-BCT.}
\label{tab:fluxes}
\begin{small}
\centering
\begin{tabular}{c@{\hspace{0.6cm}}c@{\hspace{0.6cm}}c@{\hspace{0.6cm}}ccl} \hline
Galaxy 		& Source & S/N & Corrected Flux & Luminosity & Comments \\
       		&        &     & (ergs s$^{-1}$ cm$^{-2}$) & (ergs s$^{-1}$) &\\ [10pt]
NGC\,6503	& X1 &  3.01 & $7.86 \ti 10^{-14}$ &  $4.53 \ti 10^{38}$ &\\
		& X2 &  2.51 & $6.47 \ti 10^{-14}$ &  $3.73 \ti 10^{38}$ & Probably off-nuclear\\
		& X2 &  3.47 & $2.69 \ti 10^{-13}$ &  $1.55 \ti 10^{39}$ & Total emission (r = 30\arcsec)\\[6pt]
NGC\,6946	& X1 & 11.85 & $2.05 \ti 10^{-13}$ &  $1.34 \ti 10^{39}$ &\\
		& X2 &  2.67 & $4.19 \ti 10^{-14}$ &  $2.73 \ti 10^{38}$ &\\
		& X3 & 10.73 & $1.77 \ti 10^{-13}$ &  $1.16 \ti 10^{39}$ & Nuclear\\
		& X3 & 12.80 & $4.49 \ti 10^{-13}$ &  $2.92 \ti 10^{39}$ & Total nuclear emission (r = 50\arcsec)\\
		& X4 &  4.37 & $6.10 \ti 10^{-14}$ &  $3.98 \ti 10^{38}$ &\\
		& X5 &  3.06 & $4.59 \ti 10^{-14}$ &  $2.99 \ti 10^{38}$ &\\
		& X6 &  5.48 & $7.62 \ti 10^{-14}$ &  $4.97 \ti 10^{38}$ &\\
		& X7 & 13.25 & $2.45 \ti 10^{-13}$ &  $1.59 \ti 10^{39}$ &\\
		& X8 & 32.22 & $1.15 \ti 10^{-12}$ &  $7.53 \ti 10^{39}$ & Known SNR (see text); faint red optical\\ 
		&&&&&								counterpart\\
		& X9 &  5.13 & $7.11 \ti 10^{-14}$ &  $4.64 \ti 10^{38}$ &\\
\hline
\end{tabular}
\end{small}
\end{table*}

In this section we present an atlas of X-ray images, and tabulate all
X-ray sources found satisfying the conditions described in the
previous section, for the 29 galaxies with ROSAT HRI data found in
table \ref{tab:hriobs}.

The atlas consists of maps of isointensity contour levels overlaid on
optical images. The size of the images is $\sim 6\arcmin \times 6
\arcmin$ centered on the position of the galaxy.  Whenever an optical
JKT image was not available, a $6\arcmin \times 6 \arcmin$ Digital Sky
Survey plate was used. The atlas is ordered by increasing values of
right ascension.

All detected X-ray point sources fluxes are given in table
\ref{tab:fluxes}. The sources are ordered by increasing
values of right ascension.  For each source in table \ref{tab:fluxes}
successive columns list the measured signal to noise ratio, the flux
corrected for Galactic absorption in the 0.1--2.4 keV band, and the
luminosity found assuming the distances shown in table
\ref{tab:allobs}. Comments on particular sources are given in the
last column (nuclear sources are also noted in this way). When there
is evidence of an extended component, the source is listed twice: the
flux measured in a 10\arcsec\ radius aperture is followed by the total
flux observed in a larger aperture (the radius of the second aperture
is given in the comments).

Upper limits ($2\sigma$) were obtained for galaxies without detected
nuclear sources using a 10\arcsec\ aperture located at the nuclear
positions and can be found in table \ref{tab:upperlim}.

In addition to the galaxies listed in table \ref{tab:hriobs} the
results for four galaxies with comprehensive studies of their X-ray
emission from HRI observations have been obtained directly from the
literature. Luminosities for the detected point sources in these
galaxies (and located within a $\sim 6\arcmin \times 6 \arcmin$ region
centered on the nuclei) are shown in table \ref{tab:literature}.
These values will be used during the analysis in chapter 6.

\begin{table*}
\caption{ROSAT HRI nuclear upper limits}
\label{tab:upperlim}
\centering
\begin{tabular}{@{\hspace{0.4cm}}l@{\hspace{0.6cm}}cc} \hline
Galaxy & Corrected Upper Limit & Log luminosity \\ 
       & (ergs s$^{-1}$ cm$^{-2}$) & (ergs s$^{-1}$) \\ [10pt]
NGC\,147 &    $6.05\times10^{-14}$ &      36.49\\
NGC\,185 &    $5.28\times10^{-14}$ &      36.43\\  
NGC\,205 &    $6.80\times10^{-14}$ &      36.54\\  
NGC\,221 &    $5.59\times10^{-14}$ &      36.45\\  
NGC\,247 &    $2.38\times10^{-14}$ &      37.59\\  
NGC\,1560 &   $6.00\times10^{-14}$ &      38.13\\  
NGC\,2366 &   $2.92\times10^{-14}$ &      38.14\\  
NGC\,2403 &   $4.73\times10^{-14}$ &      38.41\\  
NGC\,2976 &   $4.56\times10^{-14}$ &      37.87\\  
A\,0951+68&   $3.30\times10^{-14}$ &      36.96\\  
UGC\,6456 &   $5.79\times10^{-14}$ &      37.60\\  
NGC\,3738 &   $5.13\times10^{-14}$ &      38.36\\  
NGC\,4136 &   $2.12\times10^{-13}$ &      39.09\\  
NGC\,4144 &   $5.96\times10^{-14}$ &      38.39\\  
NGC\,4150 &   $1.51\times10^{-13}$ &      38.34\\  
NGC\,4236 &   $7.51\times10^{-14}$ &      37.98\\  
NGC\,4244 &   $9.35\times10^{-14}$ &      38.33\\  
NGC\,4605 &   $3.66\times10^{-13}$ &      39.15\\  
NGC\,5204 &   $4.53\times10^{-14}$ &      38.39\\  
NGC\,5238 &   $3.45\times10^{-14}$ &      38.27\\  
NGC\,6503 &   $5.19\times10^{-14}$ &      38.48\\
\hline
\end{tabular}
\end{table*}

\begin{table*}
\caption{ROSAT HRI sources and their luminosities (in ergs s$^{-1}$)
obtained directly from the literature for four well studied galaxies.
References: 1 \protect\scite{primini} (for Einstein HRI see
\protect\pcite{trinchieri+fabbiano}); 2:
\protect\scite{vogler+pietsch} (for Einstein HRI see
\protect\pcite{fabbiano+trinchieri2}); 3:
\protect\scite{schulman+bregman}; 4: \protect\scite{roberts+warwick}
(for Einstein HRI see \protect\pcite{fabbiano}). The superscript $a$
denotes nuclear sources; $b$ total nuclear emission estimated from the
number of counts detected within a radius r $\protect\la 1\arcmin$
\protect\cite{vogler+pietsch} after reducing the contribution from a
nearby point source (X33: source 8 in
\protect\pcite{fabbiano+trinchieri2}); $c$ total nuclear emission
measured using r=3\arcmin. No corrections were introduced to the X-ray
luminosities to correct for different adopted parameters (such as
galaxy distance, spectral models, NH column) between this paper and
the references cited above.}
\label{tab:literature}
\centering
\begin{tabular}{lclclclc} \hline
\multicolumn{2}{c}{NGC\,224$^{1}$} & \multicolumn{2}{c}{NGC\,253$^{2}$} & \multicolumn{2}{c}{NGC\,598$^{3}$} & \multicolumn{2}{c}{NGC\,3031$^{4}$}\\
source & log L & source & log L & source & log L & source & log L\\ [10pt]
27&	36.46 &	23&	37.36 & 14 &		36.91 	& 4&	37.78\\
31&	37.27 &	25&	37.36 & 15 &		36.89 	& 5&	37.38\\
32&	36.41 &	26&	36.85 & 16$^{a}$&	39.04 	& 7&	37.26\\
35&	38.25 &	28&	37.26 & 16$^{c}$&	39.11 	& 9& 	38.22\\
36&	36.63 &	29&	36.90 & & 			&10& 	37.24\\
37&	37.49 &	32&	37.04 & &			&12& 	37.24\\
38&	36.23 &	33&	38.47 & &			&13$^{a}$&39.64\\
39&	36.84 &	34$^{a}$&38.82& &			&14& 	37.41\\
40&	36.44 &	34$^{b}$&39.51 & &			&16& 	37.56\\
41&	36.92 &	35&	37.32 & & 			& & 	\\
42&	36.88 &	36&	37.90 & &			& & 	\\
43&	37.11 &	40&	37.97 & & 			& & 	\\
44$^{a}$&37.32& 42&	37.43 & & 			& & 	\\
46&	36.25 &	44& 	36.95 & & 			& & 	\\
47&	37.35 &	& & & & & \\
48&	37.20 & & & & & & \\
49&	36.83 &	& & & & & \\
50&	37.30 &	& & & & & \\
52&	37.29 &	& & & & & \\
54&	37.58 &	& & & & & \\
55&	36.50 &	& & & & & \\
57&	37.54 &	& & & & & \\
60&	37.35 &	& & & & & \\
65&	36.62 & & & & & & \\
67&	37.00 & & & & & & \\
\hline
\end{tabular}
\end{table*}

\clearpage

\begin{figure}
\centering
\caption{NGC\,147 X-ray contours overlaid on optical JKT image. 
ROSAT exposure time = 14616.6 seconds.}
\label{fig:n147xcon}
\end{figure}

\begin{figure}
\centering
\caption{NGC\,185 X-ray contours overlaid on optical JKT image. 
ROSAT exposure time = 20881.3 seconds.}
\label{fig:n185xcon}
\end{figure}

\begin{figure}
\centering
\caption{NGC\,205 X-ray contours overlaid on optical JKT image. 
ROSAT exposure time = 27842.5 seconds.}
\label{fig:n205xcon}
\end{figure}

\begin{figure}
\centering 
\caption{NGC\,221 X-ray contours overlaid on optical JKT image. 
ROSAT exposure time = 12533.5 seconds.}
\label{fig:n221xcon}
\end{figure}

\begin{figure}
\centering
\caption{NGC\,247 X-ray contours overlaid on optical JKT image. 
ROSAT exposure time = 51382.9 seconds.}
\label{fig:n247xcon}
\end{figure}

\begin{figure}
\centering
\caption{NGC\,404 X-ray contours overlaid on optical JKT image. 
ROSAT exposure time = 23562.6 seconds.}
\label{fig:n404xcon}
\end{figure}

\begin{figure}
\centering
\caption{IC\,342 X-ray contours overlaid on optical JKT image. 
ROSAT exposure time = 18990.4 seconds.}
\label{fig:ic342xcon}
\end{figure}

\begin{figure}
\centering
\caption{NGC\,1560 X-ray contours overlaid on optical Oschin Schmidt 
Telescope on Palomar Mountain as scanned for the STScI `Digital Sky
Survey'. ROSAT exposure time = 17287.5 seconds.}
\label{fig:n1560xcon}
\end{figure}

\begin{figure}
\centering
\caption{NGC\,2366 X-ray contours overlaid on optical JKT image. 
ROSAT exposure time = 31465.5 seconds.}
\label{fig:n2366xcon}
\end{figure}

\begin{figure}
\centering
\caption{NGC\,2403 X-ray contours overlaid on optical JKT image. 
ROSAT exposure time = 26244.7 seconds.}
\label{fig:n2403xcon}
\end{figure}

\begin{figure}
\centering
\caption{NGC\,2976 X-ray contours overlaid on optical JKT image. 
ROSAT exposure time = 49482.3 seconds.}
\label{fig:n2976xcon}
\end{figure}

\begin{figure}
\centering
\caption{A\,0951+68 X-ray contours overlaid on optical JKT image. 
ROSAT exposure time = 14240.2 seconds.}
\label{fig:a0951xcon}
\end{figure}

\begin{figure}
\centering
\caption{UGC\,6456 X-ray contours overlaid on optical JKT image. 
ROSAT exposure time = 13027.7 seconds.}
\label{fig:u6456xcon}
\end{figure}

\begin{figure}
\centering
\caption{NGC\,3738 X-ray contours overlaid on optical JKT image. 
ROSAT exposure time = 13711.2 seconds.}
\label{fig:n3738xcon}
\end{figure}

\begin{figure}
\centering
\caption{NGC\,4136 X-ray contours overlaid on optical JKT image. 
ROSAT exposure time = 1951.0 seconds.}
\label{fig:n4136xcon}
\end{figure}

\begin{figure}
\centering
\caption{NGC\,4144 X-ray contours overlaid on optical JKT image. 
ROSAT exposure time = 11083.4 seconds.}
\label{fig:n4144xcon}
\end{figure}

\begin{figure}
\centering
\caption{NGC\,4150 X-ray contours overlaid on optical JKT image. 
ROSAT exposure time = 14083.9 seconds.}
\label{fig:n4150xcon}
\end{figure}

\begin{figure}
\centering
\caption{NGC\,4236 X-ray contours overlaid on optical JKT image. 
ROSAT exposure time = 10933.8 seconds.}
\label{fig:n4236xcon}
\end{figure}

\clearpage

\begin{figure}
\centering
\caption{NGC\,4244 X-ray contours overlaid on optical JKT image. 
ROSAT exposure time = 8630.9 seconds.}
\label{fig:n4244xcon}
\end{figure}

\begin{figure}
\centering
\caption{NGC\,4395 X-ray contours overlaid on optical Oschin Schmidt
Telescope on Palomar Mountain as scanned for the STScI `Digital Sky
Survey'.  ROSAT exposure time = 11252.5 seconds.}
\label{fig:n4395xcon}
\end{figure}

\begin{figure}
\centering
\caption{NGC\,4605 X-ray contours overlaid on optical JKT image. 
ROSAT exposure time = 2158.4 seconds.}
\label{fig:n4605xcon}
\end{figure}

\begin{figure}
\centering
\caption{NGC\,4736 X-ray contours overlaid on optical JKT image. 
ROSAT exposure time = 138904.4 seconds.}
\label{fig:n4736xcon}
\end{figure}

\begin{figure}
\centering
\caption{NGC\,4826 X-ray contours overlaid on optical JKT image. 
ROSAT exposure time = 24633.9 seconds.}
\label{fig:n4826xcon}
\end{figure}

\begin{figure}
\centering
\caption{NGC\,5204 X-ray contours overlaid on optical JKT image. 
ROSAT exposure time = 28237.1 seconds.}
\label{fig:n5204xcon}
\end{figure}

\begin{figure}
\centering
\caption{NGC\,5236 X-ray contours overlaid on optical JKT image. 
ROSAT exposure time = 47237.0 seconds.}
\label{fig:n5236xcon}
\end{figure}

\begin{figure}
\centering
\caption{NGC\,5238 X-ray contours overlaid on optical JKT image. 
ROSAT exposure time = 23163.1 seconds.}
\label{fig:n5238xcon}
\end{figure}

\begin{figure}
\centering
\caption{NGC\,5457 X-ray contours overlaid on optical JKT image. 
ROSAT exposure time = 50813.9 seconds.}
\label{fig:n5457xcon}
\end{figure}

\begin{figure}
\centering
\caption{NGC\,6503 X-ray contours overlaid on optical JKT image. 
ROSAT exposure time = 14640.6 seconds.}
\label{fig:n6503xcon}
\end{figure}

\begin{figure}
\centering
\caption{NGC\,6946 X-ray contours overlaid on optical JKT image. 
ROSAT exposure time = 81399.7 seconds.}
\label{fig:n6946xcon}
\end{figure}

\clearpage

\section{Notes on individual objects}

In this section we briefly review the results on all 34 objects with
high resolution X-ray data. This includes objects for which ROSAT HRI
data is analyzed in this paper (tables \ref{tab:fluxes} and
\ref{tab:upperlim}), objects for which we have used ROSAT HRI results
from the literature (table \ref{tab:literature}), and Leo\,B, which
was observed with the Einstein HRI but has not been observed by the
ROSAT HRI.

{\bf NGC\,147}: NGC\,147 was observed for the first time in X-rays
using the ROSAT HRI and no emission was detected. These observations
have also been reported by \scite{brandt}. They give a $2\sigma$ upper
limit for the flux from a point source located in the galaxy of $6
\times 10^{-14}$ ergs s$^{-1}$ cm$^{-2}$ in the 0.1--2.5 keV
band-pass, which is in good agreement with the upper limit reported in
table
\ref{tab:upperlim}.

{\bf NGC\,185}: As with NGC\,147, this galaxy was observed for the
first time in X-rays and no emission was detected. These
observations have also been reported by \scite{brandt}. A $2\sigma$
upper limit for the flux of a point source was found to be $4 \times
10^{-14}$ ergs s$^{-1}$ cm$^{-2}$ in the 0.1--2.5 keV band-pass, again
in good agreement with the value quoted in table \ref{tab:upperlim}.

{\bf NGC\,205 - M\,110}: This small elliptical galaxy was observed
with the Einstein HRI and no X-ray sources were detected with a flux
$> 1.8 \times 10^{-12}$ ergs s$^{-1}$ cm$^{-2}$ in the 0.5--4.0 keV
band-pass \cite{fabbiano+etal,markert}. The ROSAT HRI observations of
NGC\,205 reported here give an upper limit of $6.78 \times 10^{-14}$
ergs s$^{-1}$ cm$^{-2}$ for the flux of any point source in the
nuclear region. The only source seen in figure \ref{fig:n205xcon} lies
2.7\arcmin\ away from the nucleus and is probably not associated with
the galaxy.

{\bf NGC\,221 - M\,32}: A strong off-nuclear source in NGC\,221 has
been found with the HRI on both Einstein and ROSAT. The flux reported
here is consistent with the Einstein measurements ($F_{X} = 9.1 \ti
10^{-13}$ ergs s$^{-1}$ cm$^{-2}$ in the 0.5--4.0 keV band-pass
\cite{fabbiano+etal}). The X-ray source lies $\sim 7\arcsec$ away 
from the NGC\,221 nucleus and has no optical counterpart. ASCA
observations of this source reported by \scite{loewenstein} reveal a
hard spectrum and a flux decrease of 25 percent in two weeks, favoring
the identification of the source as a single XRB. Variability by a
factor 3 has also been detected during ROSAT PSPC observations
\cite{supper}. An upper limit for the nuclear X-ray emission is given
in table \ref{tab:upperlim}. The position of the aperture used to
compute the upper limit was shifted slightly from the exact position
of the galactic nucleus to avoid contamination from the off-nuclear
source.

Dynamical studies of the stellar rotation velocities in this galaxy
have revealed the presence of a central dark massive object, probably
a black hole, with mass $3 \times 10^{6} M_{\odot}$
\cite{bender,vandermarel+etal}, which corresponds to an Eddington 
luminosity of $\sim 10^{44}$ ergs s$^{-1}$. The X-ray upper limit in
table \ref{tab:upperlim} shows that the central object is emitting at
most at $\la 10^{-8} L_{\rm Edd}$.

{\bf NGC\,224 - M\,31}: No analysis of X-ray data for this galaxy has
been done in this paper. Einstein and ROSAT HRI observations of M\,31
have detected over 100 individual sources with luminosities $10^{36}
\la L_{x} \la 10^{38}$ ergs s$^{-1}$ \cite{primini,trinchieri+fabbiano}. 
The ROSAT PSPC, which is a more sensitive than the HRI (but with worse
spatial resolution) gives an upper limit for a diffuse component
associated with the galactic bulge of $2.6 \times 10^{38}$ ergs
s$^{-1}$ \cite{supper}.

The point sources are associated with two components, the disk and the
bulge, and are strongly concentrated towards the centre. PSPC data
show that the bulge and disk components account for about one and two
thirds of the total emission, respectively \cite{supper}. Ginga (2--20
keV) and BeppoSAX (0.1--10 keV)  observations have shown that
this emission from the galaxy is consistent with a population of LMXRB
\cite{makishima,trinchieri+etal3}. A list of the sources found within
$\sim 6\arcmin$ from the nucleus is shown in table
\ref{tab:literature}. The source coincident with the galactic nucleus 
has been reported to vary \cite{primini} and is probably an XRB.

{\bf NGC\,247}: Two strong X-ray sources are seen in the southern
region of the ROSAT HRI image (see figure \ref{fig:n247xcon}). A
$2\sigma$ upper limit for a point source located in the nuclear region
of the galaxy can be found in table \ref{tab:upperlim}.  \scite{read}
report the detection of 5 ROSAT PSPC sources associated with the
galaxy, 3 of which lie outside our optical image. The remaining two
correspond to our sources X1 and X2. No obvious optical couterparts
are seen at these positions, although X1 is located at the edge of a
bright star-forming region. Notice that the astrometry of the
observations is thought to be accurate (section 3.2.2).

From a fit in the 0.1--2.0 keV range, \scite{read} found that the PSPC
spectrum of X1 indicates a very soft and obscured source ($kT = 0.12$
keV, $N_{H} = 6.4 \times 10^{21}$ cm$^{-2}$). Using their spectral
parameters we find that the HRI count rate corresponds to a flux in
the 0.1--2.0 keV energy range, (corrected by Galactic absorption only)
of $5.0 \times 10^{-13}$ ergs s$^{-1}$ cm$^{-2}$, compared with the
PSPC flux of $2.1 \times 10^{-13}$ ergs s$^{-1}$ cm$^{-2}$ given by
\scite{read}. This corresponds to an increase in the source luminosity
by more than a factor 2 between the PSPC and HRI observations. From
the HRI count rate, the intrinsic luminosity of the source is found to
be $\sim 10^{42}$ ergs s$^{-1}$ for an assumed distance to NGC\,247 of
3.69 Mpc, although variations in the adopted parameters can change
this significantly.  For example, changing the hydrogen column and
temperature by 1$\sigma$ each (using the error bars in \scite{read},
ie., $kT = 0.15$ keV, $N_{H} = 4.7 \times 10^{21}$ cm$^{-2}$)
decreases the luminosity by an order of magnitude.  The very soft
spectral distribution, extremely high intrinsic luminosity, and
detected variability make this source a good candidate for a stellar
accreting black hole.

\scite{mackie+etal} also report PSPC observations and find a
faint nuclear source with $L_{X} = 1 \times 10^{36}$ ergs s$^{-1}$,
which is well below the detection limit of our HRI
observations. However, due to the poor spatial resolution of the PSPC,
it is not clear whether this source is coincident with the galaxy
nucleus.

{\bf NGC\,253}: No analysis of X-ray data for this galaxy has been
done in this paper. This starburst galaxy has very complex X-ray
emission. Thirty one point sources have been detected with the
Einstein and ROSAT HRI \cite{vogler+pietsch} with luminosities of up
to a few times $10^{38}$ ergs s$^{-1}$ (see table \ref{tab:literature}). 
Analysis of HRI and PSPC observations show that the source located in
the nuclear region is extended, soft and possibly variable
\cite{vogler+pietsch}. Spectral fits to some disk sources using PSPC 
observations show that they are consistent with the spectra of
absorbed XRBs \cite{read}.

Extended emission is detected well above and below the galactic plane
of this galaxy (see figure 11 and 12 in \pcite{dahlem+etal}) and
NGC\,253 is a prototype object for the study of X-ray emission from
starburst galaxies. Combined ASCA (FWHM $\sim 3\arcmin$) and PSPC
observations show that three components are required to fit the
integral spectrum of the galaxy: a power law ($\Gamma \sim 1.9$) and a
two-temperatur plasma ($kT \sim 0.3$ keV and $kT \sim 0.7$)
\cite{dahlem+etal}. The determination of more detailed physical
parameters is hampered by the difficulty of measuring multiple
absorbing columns, as discussed by \scite{dahlem+etal}. Their analysis
of the halo diffuse emission detected with the PSPC shows that the
spectrum is well fitted by a two-temperatur plasma ($kT \sim 0.1$ keV
and $kT \sim 0.7$) if solar abundances are assumed. BeppoSAX
observations of NGC\,253 show the presence of a $\sim 300$ eV Fe\,K
line at 6.7 KeV, probably of thermal origin \cite{persic}. Other
prominent lines are also resolved with ASCA \cite{ptak+etal}.

{\bf NGC\,404}: A weak X-ray nuclear source ($F_{X} = 7.5 \times
10^{-12}$ ergs s$^{-1}$) has been detected by the ROSAT HRI in the
LINER nucleus of this galaxy. An ASCA 2--10 keV upper limit of $3
\times 10^{-13}$ ergs s$^{-1}$ cm$^{-2}$ has been reported by
\scite{maoz+etal}, which implies that the ROSAT flux is either dominated
by very soft emission, or that the source is variable. The former idea
is supported by recent UV observations of the NGC\,404 nucleus which
show that the spectrum is dominated by stellar absorption features
from massive young stars \cite{maoz+etal} and not by the blue,
featureless continuum expected from an active nucleus.

{\bf NGC\,598 - M\,33}: No analysis of X-ray data for this galaxy has
been done in this paper. Several point sources have been detected with
the Einstein and ROSAT HRI in this spiral galaxy
\cite{markert+rallis,schulman+bregman}. Those sources located within
the central $\sim 6\arcmin \times 6\arcmin$ region are listed in table
\ref{tab:literature}. The most striking source is the nucleus, with a
luminosity of $\ga 10^{39}$ ergs s$^{-1}$ which makes it a good AGN
candidate. However, the nucleus is not detected at radio wavelengths
and shows very little line emission in the optical (Schulman \&
Bregman 1995, Paper III). Nuclear dynamical studies also give a strict
limit for a central black hole mass of $\la 5 \times 10^{4} M_{\odot}$
\cite{kormendy+mcclure}. ASCA observations show that the X-ray
emission from this source is much softer than the typical AGN spectrum
and it does not show signs of variability to within 10\% on time
scales of 100 minutes \cite{takano}. This almost certainly excludes
the possibility of an active nucleus in NGC\,598. In fact, a disk
blackbody fit to the ASCA data shows that its emission is similar to
Galactic black hole candidates \cite{takano,colbert}. Diffuse emission
around the nucleus has been detected in ROSAT HRI as well as PSPC
observations \cite{schulman+bregman,long+etal,read}.

{\bf IC\,342}: We have re-analyzed the ROSAT HRI data already
discussed by \scite{bregman}. They found evidence for a diffuse nuclear
component that could be explained as a hot interstellar medium
generated by a very young nuclear starburst.

\begin{figure}
\centering
\includegraphics[bb=60 70 570 670,angle=270,scale=0.4]{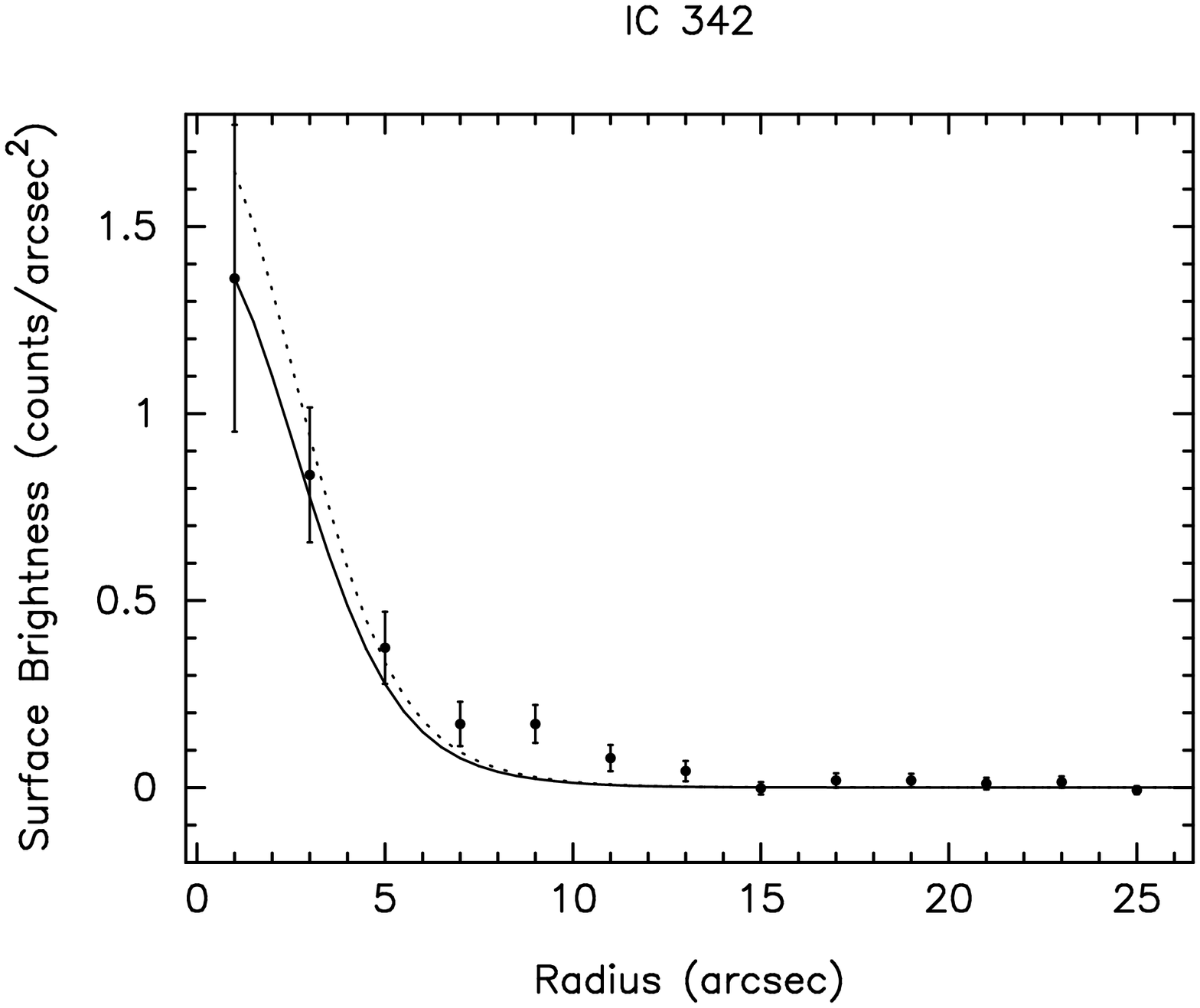}
\includegraphics[bb=60 70 570 670,angle=270,scale=0.4]{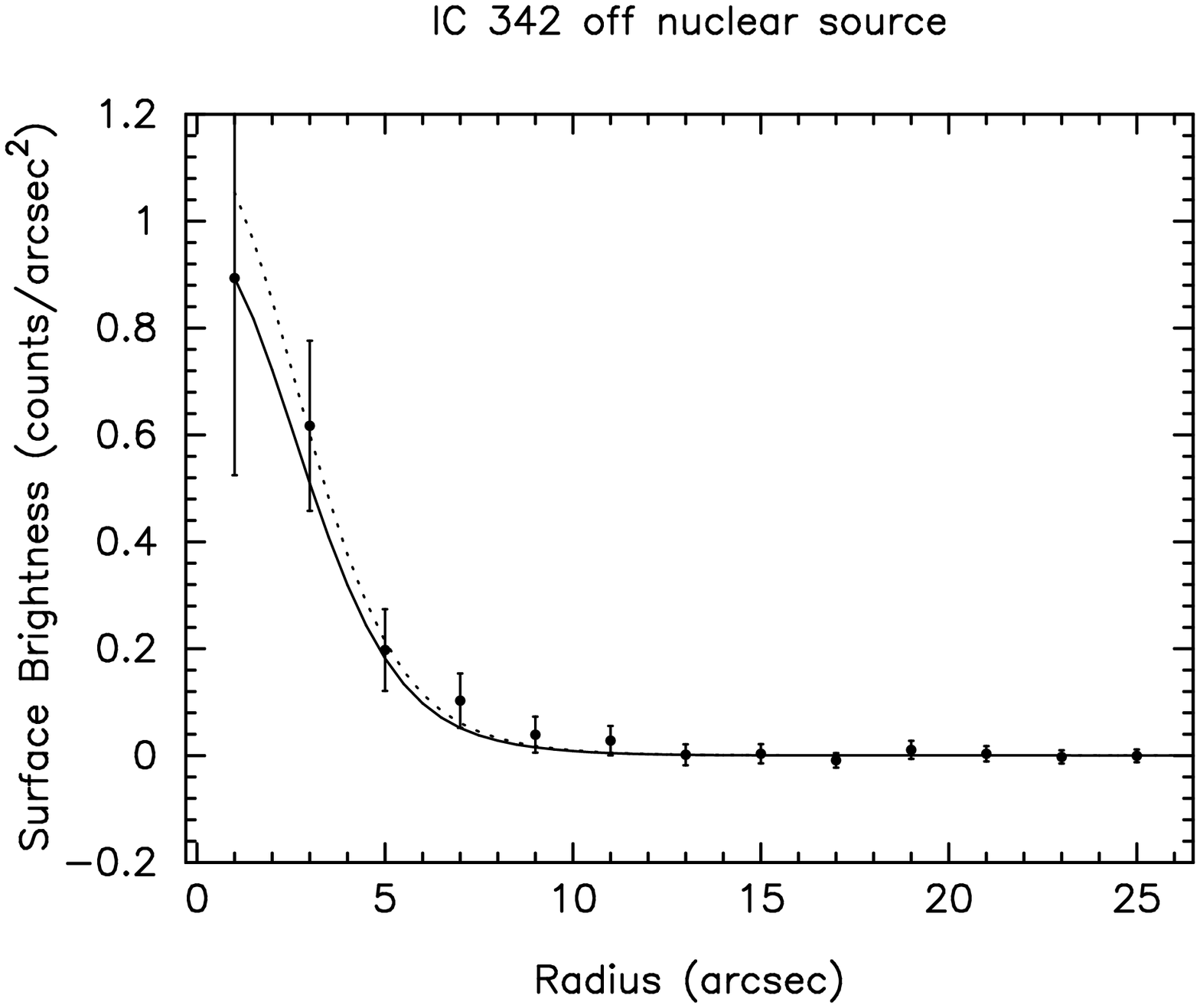}
\caption{Observed profile and model PSF for IC\,342 X-2 and an
off-nuclear source. The solid line corresponds to a model PSF
\protect\cite{david+etal} scaled to match the peak of the observed
profile. The dashed line corresponds to the same model PSF but with a
scale factor found by minimizing a chi-squared fit to the data.}
\label{fig:psf.ic342}
\end{figure}

Figure \ref{fig:ic342xcon} shows that three sources lie within the
limits of the optical JKT image of the galaxy. The brightest source
is coincident with the nucleus and visual inspection of the image
suggests that it is marginally resolved. A comparison of the
azimuthally averaged profile of the source and the HRI model PSF can
be seen in figure \ref{fig:psf.ic342} and a clear deviation from the
model PSF is visible at a radius of between 7 and 13 arcsec
from the centre. For comparison, an off-nuclear source is also shown,
which is in good agreement with the model PSF, confirming the diffuse
nuclear component.

X3 is located close to a faint knot of optical emission while X1 has
no obvious optical counterpart. Notice however, that only one source
was used to check the astrometry of this observation (section
3.2.2). X1 has a S/N of 2.3 (see table \ref{tab:fluxes}), but
is included here because it was reported by \scite{bregman} (they
found a S/N of 2.8, probably due to a different background estimate).

\scite{fabbiano+trinchieri} analyzed Einstein IPC observations of the
nuclear region in IC\,342 and argued that the emission is consistent
with starburst activity.  Unfortunately, the IPC was not able to
resolve the three sources detected with the HRI. 

{\bf NGC\,1560}: This galaxy was observed for the first time in X-rays
and no emission was detected in the ROSAT HRI data.  A $2\sigma$ upper
limit for a point source located in the nuclear region of the galaxy
can be found in table \ref{tab:upperlim}.

{\bf NGC\,2366}: No emission was detected in the ROSAT HRI
observations of this galaxy.  A $2\sigma$ upper limit for a point
source located in the nuclear region of the galaxy can be found in
table \ref{tab:upperlim}.

{\bf NGC\,2403}: This galaxy was observed with the Einstein HRI and
IPC instruments \cite{fabbiano+trinchieri}. Three prominent sources
were identified in the IPC observations with no emission from the
nuclear region of the galaxy.

The ROSAT HRI observations reported here show a total of 4 point
sources associated with the galaxy. Sources X1 and X3 in figure
\ref{fig:n2403xcon} correspond to two of the sources reported by
\scite{fabbiano+trinchieri} (their third source lies outside the JKT
image). Sources X2 and X4 are about one order of magnitude fainter
than X1 and X3, and were probably below the sensitivity threshold of
the Einstein observations. It is confirmed that no nuclear source is
present in the galaxy and a $2\sigma$ upper limit for a nuclear point
source is given in table \ref{tab:upperlim}.

A search for SNRs in NGC\,2403 has yielded 35 detections
\cite{matonick}. The position of remnant number 15 in table 2 of
\scite{matonick} is coincident with the X-ray source X3 reported here
(see table \ref{tab:fluxes}). A very faint optical counterpart is
observed at this position. If the identification is correct the SNR
would belong to a class of super-luminous (probably young) remnants
\cite{schlegel3}.

Source X4 is coincident with a giant HII region. A photometric study
of this region (N2403-A) reveals more than 1400 detected stars, among
them 800 O-type stars and a lower limit of 23 WR stars
\cite{drissen+roy}.

The coincidence of X3 and X4 with previously known sources in NGC\,2403
gives support to the astrometric checks carried out for this galaxy
(section 3.2.2).

\begin{figure}
\centering
\setlength{\fboxrule}{1pt}
\setlength{\fboxsep}{0cm}
\fbox{\includegraphics[bb=110 245 500 638,scale=0.5]{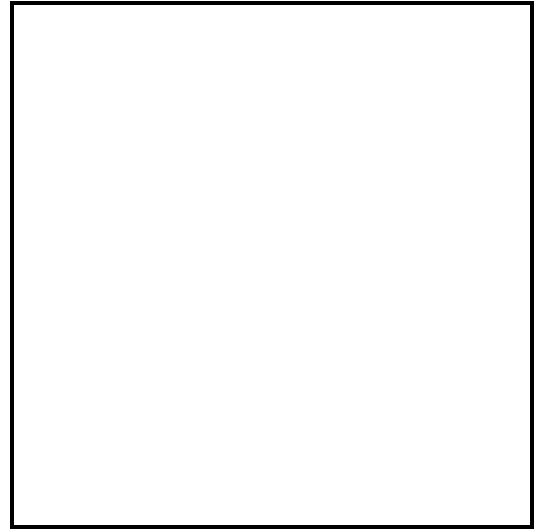}}
\caption{Optical I-band image of UGC\,6456. The position of the
optical counterpart for the X-ray source X1 is indicated.}
\label{fig:u6456.i}
\end{figure}

{\bf NGC\,2976}: This galaxy was observed for the first time in X-rays
and no emission was detected in the ROSAT HRI data. A $2\sigma$ upper
limit for a point source located in the nuclear region of the galaxy
can be found in table \ref{tab:upperlim}.

The only detected X-ray source has no obvious counterpart. Notice,
however, that the astrometry of the observations is highly uncertain
given that only one optical identification was found in the field of
view of the HRI and a significant shift was applied to one of the
images (section 3.2.2).

{\bf A\,0951+68}: This galaxy has been observed in X-rays for the
first time. As can be seen in figure \ref{fig:a0951xcon}, the only
detected source is probably a foreground or background object.

{\bf NGC\,3031}: No analysis of X-ray data for this galaxy has been
done in this paper. Nine and 30 point sources have been detected in
this galaxy from observations with the Einstein and ROSAT HRI, respectively
\cite{fabbiano,roberts+warwick}. All Einstein sources within the
galaxy D$_{25}$ isophote were detected by ROSAT with the exception of
the Einstein source X1. ROSAT sources located within the central $\sim
6\arcmin \times 6\arcmin$ region are listed in table
\ref{tab:literature}.

The central emission is dominated by the nuclear source, coincident
with a low-luminosity active nucleus, with a luminosity $\la 10^{40}$
ergs s$^{-1}$ in the 0.2--4.0 keV band-pass. Einstein IPC data show
that the spectrum of the nuclear source is soft, with a good fit given
by thermal emission with $kT \sim 1$ keV or by a power law with index
$\alpha \sim 2$ \cite{fabbiano}. Broad band observations obtained with
BBXRT (0.5--10 keV) and ASCA (0.2--10 keV) show, however, that the
nuclear emission is consistent with a power law distribution with
index $\alpha \sim 1$
\cite{petre+etal,ishisaki,serlemitsos+ptak+yaqoob}.  The ASCA data
also revealed the presence of a broad iron K emission line similar to
those seen in more luminous Seyfert galaxies
\cite{ishisaki,serlemitsos+ptak+yaqoob}. It must be kept in mind that
due to the coarse spatial resolution of ASCA some contamination is
expected from nearby sources, although \scite{ishisaki} estimated this
to be less than 10 percent. X-ray long term and fast variability by
significant factors have been reported for the nuclear source
\cite{petre+etal,ishisaki,serlemitsos+ptak+yaqoob}.

{\bf Leo\,B}: No analysis of X-ray data for this galaxy has been done
in this paper. Leo\,B was observed with the Einstein
HRI. \scite{markert} report that no sources were detected.

\begin{figure}
\centering
\includegraphics[scale=0.6]{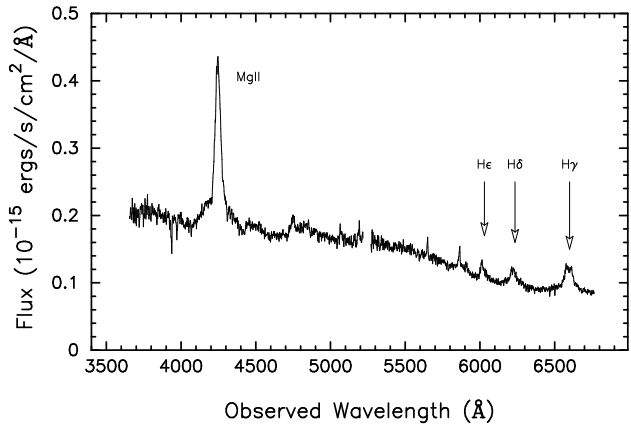}
\caption{Background quasar in NGC\,4150.}
\label{n4150_qso}
\end{figure}

\begin{figure}
\centering
\setlength{\fboxrule}{1pt}
\setlength{\fboxsep}{0cm}
\fbox{\includegraphics[bb=103 239 508 645,scale=0.5]{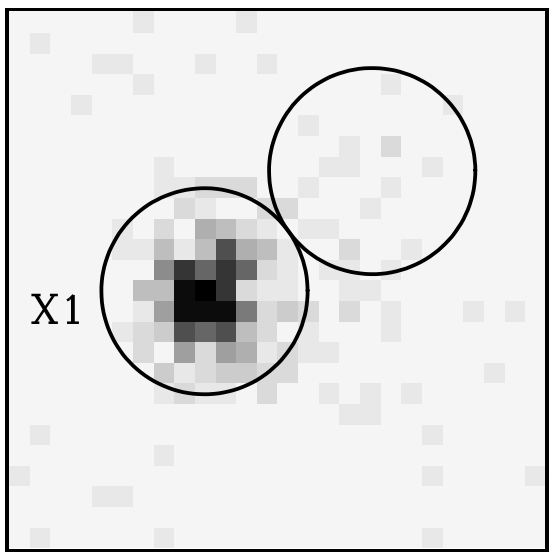}}
\caption{Raw X-ray image of NGC\,4150 (pixel size $= 2\arcsec \times
2\arcsec$) showing the location of the quasar and nucleus apertures.}
\label{fig:n4051.ap}
\end{figure}

{\bf UGC\,6456}: This is a blue compact galaxy. These galaxies are
characterized by low metallicities, high gas content and vigorous star
formation. In UGC\,6456, evidence of both a recent episode of strong
star formation (600--700 Myr) and an older stellar population has been
found \cite{lynds}. 

The JKT optical images show that the galaxy has numerous bright knots
of emission surrounded by a low surface brightness outer envelope
(Johnson 1997; Paper II). The knots of emission are displaced south
from the geometrical center of the envelope and it is not clear
whether they represent the true nuclear region of the galaxy (figure
\ref{fig:u6456.i}). 

ROSAT PSPC observations of this galaxy show a central X-ray core and
three extended structures connected to the central source
\cite{papaderos}. This morphology was interpreted as outflows from the
central region of the galaxy, powered by starburst activity. The total
PSPC flux within a circular aperture of radius 3\arcmin\ is $1.6 \ti
10^{-13}$ ergs s$^{-1}$ cm$^{-2}$. 

The high resolution data reported here show a strong X-ray source
located at the north-most limit of the optical emission knots, but
displaced to the west with respect to the geometrical centre of the
outer envelope (figure \ref{fig:u6456.i}). Notice also that the
astrometry of the HRI observation is fairly well established (section
3.2.2). There is no evidence of extended X-ray emission in the
observations, probably due to the lower sensitivity of the HRI. The
HRI flux of the point source is $\sim 9 \times 10^{-13}$ ergs s$^{-1}$
cm$^{-2}$, about 5 times more luminous than the PSPC observations. A
very luminous XRB ($L_{X} \ga 10^{38}$ ergs s$^{-1}$) could be
responsible for this flux variation.

{\bf NGC\,3738}: This galaxy has an irregular optical appearance with
several bright knots of emission, although the outer parts are quite
regular.  None of the bright knots seems to coincide with the
geometrical center of the galaxy (Johnson 1997; Paper II).

NGC\,3738 has been observed in X-rays for the first time and no point
sources associated with the galaxy have been found. An upper limit for
the X-ray emission can be found in table \ref{tab:upperlim}. Source X1
in figure \ref{fig:n3738xcon} is coincident with a faint knot of
optical emission and is probably a foreground or background
object. Source X2 is coincident with a bright point-like object and
probably corresponds to a foreground star.

{\bf NGC\,4136}: This spiral galaxy has been observed for the first
time in X-rays.  No sources associated with the nuclear region have
been found.  An upper limit to the flux from a nuclear point source
can be found in table \ref{tab:upperlim}. A strong X-ray source is
coincident with one of the spiral arms of the galaxy where several
knots of emission can be seen in the optical image (see figure
\ref{fig:n4136xcon}). The remnant of the historic Type II supernova
SN\,1941C seen in NGC\,4136 is not located close to this X-ray source
\cite{vandyk}.

{\bf NGC\,4144}: This galaxy was observed for the first time in X-rays
and no emission was detected in the ROSAT HRI data. A $2\sigma$ upper
limit for a point source located in the nuclear region of the galaxy
can be found in table \ref{tab:upperlim}.

{\bf NGC\,4150}: A strong point-like source coincident with this
galaxy was detected in the ROSAT All-Sky Survey with $F_{X} = 6
\times 10^{-13}$ ergs s$^{-1}$ cm$^{-2}$ and a photon index $\Gamma =
1.41$. The emission was assumed to be from the nucleus of NGC\,4150
\cite{moran,boller}. The high resolution image seen in figure 
\ref{fig:n4150xcon} shows, however, that the X-ray source is 
more than 15\arcsec\ away from the galactic nucleus and has a position
consistent with a knot of optical emission.  Spectroscopy of the
optical counterpart shows that the source is a background quasar at
redshift 0.52 (figure \ref{n4150_qso}).

Figure \ref{fig:n4150xcon} shows that the X-ray contours of the source
are elongated in the north west direction, suggesting that {\em some}
emission might be coming from the nuclear region of the galaxy. An
estimate of the nuclear emission was obtained using a 10\arcsec\
radius aperture located as shown in figure \ref{fig:n4051.ap}.
Although the observed counts have a S/R $\sim 2.6$ (and so would be
considered a significant detection by the criteria defined in section
3.2.2) the measurement will be treated as an upper limit because of
contamination from the nearby quasar.

{\bf NGC\,4236}: This galaxy has a very low surface brightness and no
obvious nucleus (Johnson 1997; Paper II). From the ROSAT HRI
observations reported here no X-ray sources have been found in the
central region of the galaxy. The only detected source (X1) is located
in the galactic plane and might have a faint optical counterpart.
Notice, however, that the astrometry of the observations is highly
uncertain given that only one optical identification was found in the
field of view of the HRI and a significant shift was applied to both
coadded images (section 3.2.2). A $2\sigma$ upper limit for a point
source located in the nuclear region of the galaxy can be found in
table \ref{tab:upperlim}.

{\bf NGC\,4244}: No emission was detected in the ROSAT HRI
observations of this galaxy. A $2\sigma$ upper limit for a point
source located in the nuclear region of the galaxy can be found in
table \ref{tab:upperlim}.

{\bf NGC\,4395}: This galaxy contains the faintest and nearest Seyfert
1 nucleus known today \cite{filippenko+sargent,lira}. Its nuclear
X-ray source is highly variable and the continuum is well fitted by a
power-law distribution with photon index $\Gamma = 1.7$
\cite{iwasawa}. The bright source seen in figure \ref{fig:n4395xcon}
(X2) has no obvious optical counterpart.

\begin{figure}
\centering
\includegraphics[bb=60 70 570 670,angle=270,scale=0.4]{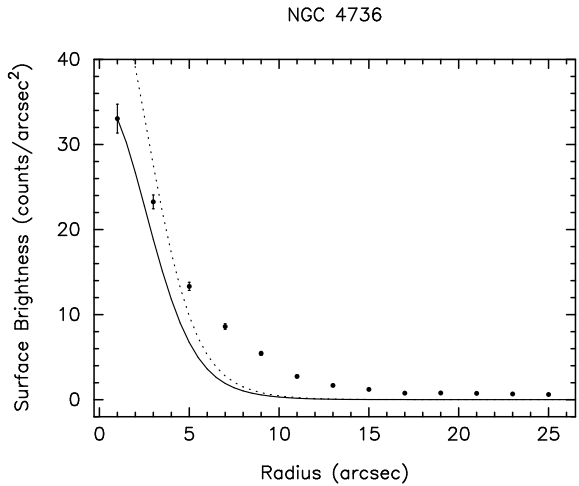}
\caption{Observed profile and model PSF for the nuclear X-ray source
in NGC\,4736. Models as in figure \protect\ref{fig:psf.ic342}.}
\label{fig:psf.n4736}
\end{figure}

\begin{figure}
\centering
\includegraphics[bb=60 70 570 670,angle=270,scale=0.4]{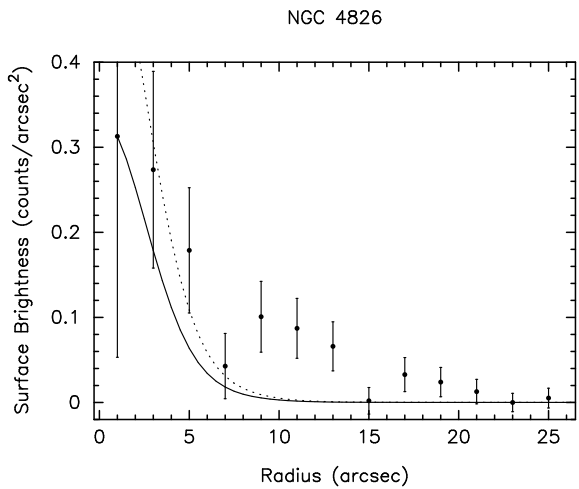}
\caption{Observed profile and model  PSF for the nuclear X-ray source 
in NGC\,4826. Models as in figure \protect\ref{fig:psf.ic342}.}
\label{fig:psf.n4826}
\end{figure}

{\bf NGC\,4605}: This galaxy was observed for the first time in X-rays
and no emission was detected in the ROSAT HRI data. A $2\sigma$ upper
limit for a point source located in the nuclear region of the galaxy
can be found in table \ref{tab:upperlim}.

{\bf NGC\,4736 - M\,94}: Strong X-ray emission is associated with the
LINER nucleus of this galaxy. An extended nuclear source can be seen
in figure \ref{fig:n4736xcon}. The azimuthally averaged profile of
this source and the HRI model PSF are shown in figure
\ref{fig:psf.n4736}.

The galaxy was previously imaged with the Einstein HRI. A nuclear flux
of $2.0\times 10^{-12}$ ergs s$^{-1}$ cm$^{-2}$ was measured within an
aperture of 60\arcsec\ radius in the 0.5--4.0 keV band-pass
\cite{fabbiano+etal}. The fluxes measured from the ROSAT HRI
observation using a small (r = 10\arcsec) and a large aperture (r =
100\arcsec) are $1.4 \times 10^{-12}$ and $4.7 \times 10^{-12}$ ergs
s$^{-1}$ cm$^{-2}$ respectively (see table \ref{tab:fluxes}). Since
the emission is highly concentrated towards the nucleus, the
difference between the 60\arcsec\ Einstein and 100\arcsec\ ROSAT
fluxes cannot be explained by the different aperture sizes alone. The
change in the total observed flux could be explained, however, if the
central source is variable or if a substantial fraction of the
emission is radiated in the very soft X-rays ($kT \la 0.5$ keV).

\scite{roberts+etal} have recently reported on ASCA and ROSAT
PSPC and HRI observations for NGC\,4736. They find that the nuclear
emission is consistent with an unresolved source plus an extended
component. PSF modeling shows that the unresolved source accounts for
more than 50\% of the detected emission \cite{roberts+etal}. The
0.1--10 keV ASCA spectrum of the emission is consistent with a
power-law with index $\alpha \sim 1$ plus a softer thermal component
($kT \sim 0.1-0.6$ keV) which dominates below 2 keV.

{\bf NGC\,4826}: The nucleus of NGC\,4826 has been classified as a
transition object (a combination of a LINER and an HIIR nucleus) by
\scite{hfs}. The galaxy was observed with the Einstein IPC and a
nuclear flux of $7.89 \times 10^{-13}$ ergs s$^{-1}$ cm$^{-2}$ was
measured within a 4.5\arcmin\ radius aperture \cite{fabbiano+etal}, in
good agreement with the flux found in table \ref{tab:fluxes}. Figure
\ref{fig:psf.n4826} shows the azimuthally averaged profile as observed
by the ROSAT HRI. Significant extended emission is observed within
$\ga 20\arcsec$ of the central peak.

{\bf NGC\,5204}: Einstein IPC observations of this galaxy show strong
X-ray emission with a total flux of $9.6 \times 10^{-13}$ ergs
s$^{-1}$ cm$^{-2}$ \cite{fabbiano+etal} in good agreement with the
measurement given here (see table \ref{tab:fluxes}). Population I XRBs
were thought to be responsible for this emission since the number of
OB stars inferred from IUE observations were in agreement with the
X-ray luminosity \cite{fabbiano+panagia}. However, the HRI image (see
figure \ref{fig:n5204xcon}) shows that the X-ray emission is
consistent with a single off-nuclear point source $\sim 17\arcsec$
away from the nucleus. Since only one optical ID was found in the
field of view of the HRI, the astrometric solution of these
observations is quite uncertain. However, the very good agreement in
the pointing of the two coadded exposures argues in favor of a fairly
accurate astrometry.

Three SNRs have been identified in NGC\,5204, but none of them is
coincident with the position of the X-ray source \cite{matonick2}.
Although there is no obvious optical counterpart for the source,
several nearby optical knots can be seen in figure
\ref{fig:n5204xcon}. The spectrum of one of the candidates, which
corresponds to a star forming region, will be presented in Paper III.

The extremely high luminosity of the X-ray source, together with the
lack of variability observed between the Einstein and the ROSAT
observations, favor an identification as a SNR from a class of
super-luminous remnants (the SASS report does not find conclusive
evidence of variability during the HRI observations, either). However,
the null detection by \scite{matonick2} argues against this
hypothesis, unless the remnant has an unusually low SII/H$\alpha$
ratio (a value $\geq 0.45$ was adopted by \scite{matonick2} as
selection criteria to identify SNRs) or unsually high
$L_{X}$/H$\alpha$ ratio (the super-luminous remnant in NGC\,6946 - see
below - has $L_{X}$/H$\alpha \sim 15$, which is the highest value seen
in these type of objects; the SNR in NGC\,5204 would require
$L_{X}$/H$\alpha > 550$).

\begin{table}
\caption{Count rates for the point  sources observed in NGC\,5236 from
two ROSAT HRI observations obtained in January 1993 and July 1994. $a$: 
nuclear source.}
\label{tab:n5236}
\centering
\begin{tabular}{|c@{\hspace{0.6cm}}c@{\hspace{0.6cm}}c@{\hspace{0.6cm}}|} \hline
Source 	& CR (20/01/93) & CR (30/07/94) \\
       	&(counts ks$^{-1}$)&(counts ks$^{-1}$) \\ [10pt]
X1 	&	$1.27 \pm 0.31$	&  $0.38 \pm 0.22$ \\
X2 	&	$1.18 \pm 0.31$	&  $1.39 \pm 0.32$ \\
X3$^{a}$ &	$15.20 \pm 0.86$&  $17.57 \pm 0.91$\\
X4 	&	$0.75 \pm 0.27$	&  $0.88 \pm 0.28$ \\
X5 	&	$0.67 \pm 0.26$	&  $0.64 \pm 0.25$ \\
X6 	&	$2.08 \pm 0.37$	&  $2.43 \pm 0.39$ \\
X7 	&	$1.91 \pm 0.36$	&  $0.05 \pm 0.19$ \\
X8 	&	$1.31 \pm 0.32$	&  $0.17 \pm 0.20$ \\
\hline 
\end{tabular}
\end{table}

\begin{figure}
\centering
\includegraphics[bb=60 70 570 670,angle=270,scale=0.4]{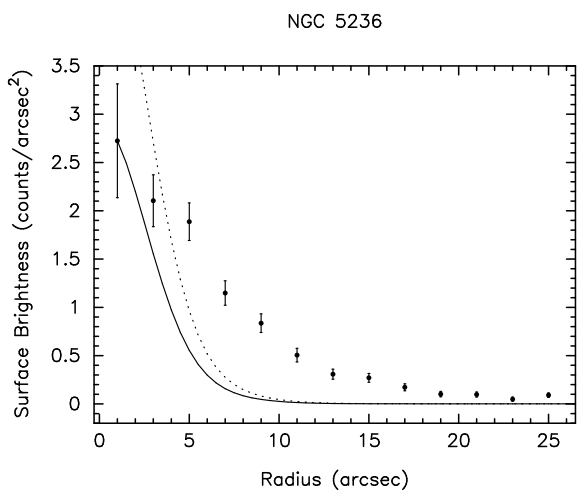}
\caption{Observed profile and model PSF for the nuclear X-ray source
in NGC\,5236. Models as in figure \protect\ref{fig:psf.ic342}.}
\label{fig:psf.n5236}
\end{figure}

{\bf NGC\,5236 - M\,83}: The very complex X-ray emission in this
galaxy can be appreciated in figure \ref{fig:n5236xcon}. Several X-ray
knots are distributed on top of bright, uneven extended
emission. Comparison between the two ROSAT HRI observations reveals
that at least three of the eight detected point sources are
variable. Table \ref{tab:n5236} shows the count rates observed in
January 1993 and July 1994. Sources X1, X7 and X8 are not detected
during the observations obtained in July 1994, but are among the
brightest objects seen in January 1993.

\begin{figure}
\centering
\includegraphics[bb=60 70 570 670,angle=270,scale=0.4]{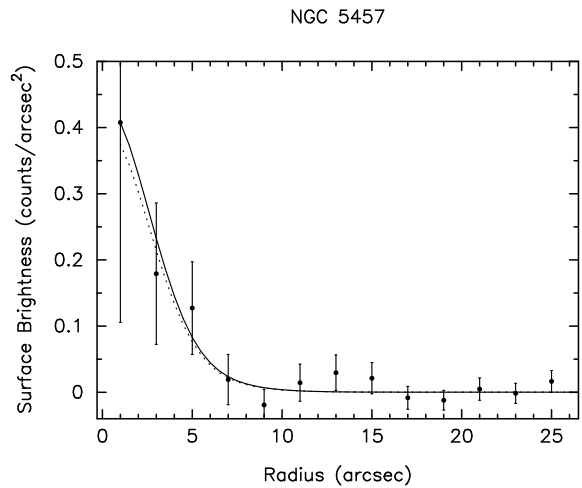}
\caption{Observed profile and model PSF for the nuclear X-ray source
in NGC\,5457. Models as in figure \protect\ref{fig:psf.ic342}.}
\label{fig:psf.n5457}
\end{figure}

\scite{trinchieri+etal} reported on the Einstein HRI
observations of this galaxy. From their observations (obtained in
January 1980 and February 1981) only 3 sources were detected in the
nuclear region of the galaxy. Two correspond to the ROSAT sources X3
(the nucleus) and X6, which are the brightest sources observed in the
ROSAT HRI data (see table \ref{tab:n5236}). Their fluxes in the
Einstein (0.5--3.0 keV) band-pass are in good agreement with the
fluxes given in table \ref{tab:fluxes}. The third source had a
luminosity of $2.3 \times 10^{38}$ ergs s$^{-1}$ in the Einstein
band-pass (for a distance to NGC\,5236 of 3.75 Mpc as assumed by
\pcite{trinchieri+etal}), and it is not detected in the ROSAT
data. It probably corresponds to a transient XRB.

\scite{ehle} report on ROSAT PSPC observations of NGC\,5236 obtained
between January 1992 and January 1993. They find a luminosity for the
nuclear source of $7 \times 10^{-13}$ ergs s$^{-1}$ cm$^{-2}$, in good
agreement with the measurement in table \ref{tab:fluxes}. They also
detect all the point sources seen in the HRI observations, with the
exception of X5, which corresponds to the faintest source in the
nuclear region. This suggests that the variable sources X1, X7 and X8
were detectable for at least a year (from the beginning of 1992 until
the beginning of 1993), before fading away, becoming undetectable by
July 1994.

The diffuse emission from the central region of NGC\,5236 can be
appreciated in figure \ref{fig:psf.n5236}.  From the PSPC observations
\scite{ehle} find that the soft 0.1-0.4 keV diffuse component accounts
for almost half of the total X-ray emission and argue that most of it
is due to hot gas in a super bubble with radius $\sim 10-15$
kpc. Evidence of vigorous starburst activity comes from observations
of the nuclear and circumnuclear regions of NGC\,5236 which have
intricate morphologies in the UV, optical and infrared
\cite{bohlin,johnson,gallais}. 

Finally, several historic supernovae have been observed in this
galaxy, but none of them is consistent with the positions of the point
X-ray sources.

{\bf NGC\,5238}:  This galaxy was observed for the first time in X-rays
and no emission was detected in the ROSAT HRI data. A $2\sigma$ upper
limit for a point source located in the nuclear region of the galaxy
can be found in table \ref{tab:upperlim}.

\begin{figure}
\setlength{\fboxrule}{1pt}
\setlength{\fboxsep}{0cm}
\centering
\fbox{\includegraphics[bb=110 248 500 638,scale=0.5]{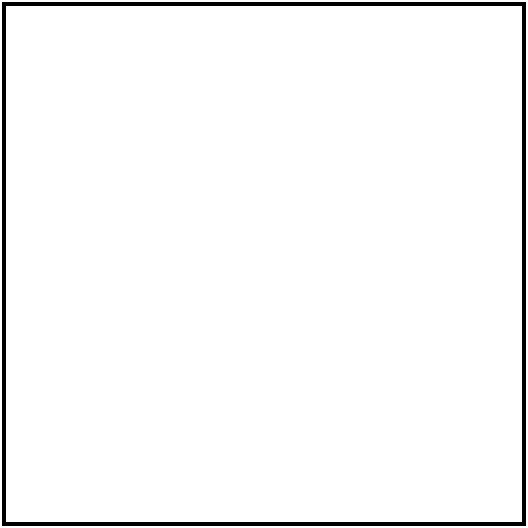}}
\caption{X-ray isocontours plotted on top of an optical JKT image of 
NGC\,6503.  The X-ray image was produced binning the raw frame into
2\arcsec\ pixels and smoothed using Gaussians with $\sigma =
8\arcsec$. Contours were drawn at $1.5^{n}$ times the standard
deviation per pixel in the smoothed background, where $n = 1, 2, 3,$
etc.}
\label{fig:n6503xcon.4}
\end{figure}

{\bf NGC\,5457 - M\,101}: Observations of this galaxy with the
Einstein IPC and ROSAT PSPC have been widely reported
\cite{trinchieri+etal2,murphy,snowden}. Results from an ultra-deep
(229 ks) ROSAT HRI observation have become available recently
\cite{wang+etal}. The data reported here were the result of combining
2 of the 4 images used to produce the ultra-deep observation reported
by \scite{wang+etal}.

\scite{wang+etal} find a total of 51 point sources down to fluxes 
$\sim 6\times 10^{-15}$ ergs s$^{-1}$ cm$^{-2}$, of which about half
are thought to be associated with the galaxy. The X-ray emission
beautifully traces the spiral arms of the galaxy and 5 of the
individual sources are associated with giant HIIRs \cite{murphy}, but
they are located outside the $\sim 6\arcmin \times 6\arcmin$ optical
JKT image shown in figure \ref{fig:n5457xcon}.

From IPC data \scite{trinchieri+etal2} find a nuclear X-ray flux of $3
\times 10^{-13}$ ergs s$^{-1}$ cm$^{-2}$ within a circle of 90\arcsec\
radius for the Einstein (0.2--4.0 keV) band-pass. From the ROSAT HRI
observations a nuclear flux of $4 \times 10^{-14}$ ergs s$^{-1}$
cm$^{-2}$ is obtained, an order of magnitude fainter than the IPC
flux. The difference can be explained if sources X2 and X3, which are
not resolved by the IPC, were contained within the large aperture used
by \scite{trinchieri+etal2}, or by the effect of luminous and highly
variable XRBs. The ROSAT HRI count rate found by \scite{wang+etal} for
the nuclear source is identical to our value (0.67 counts ks$^{-1}$),
although the inferred fluxes disagree by a factor $\sim 1.5$, probably
because different spectral models were assumed.

The presence of a soft diffuse component is discussed by
\scite{snowden} and \scite{read}. They find conclusive evidence in
ROSAT PSPC observations for extended emission within the inner
7\arcmin\ of the galaxy. The radial profile obtained from the ROSAT
HRI observations shown in figure \ref{fig:psf.n5457} suggests some
patchiness in the X-ray. The better signal to noise data analyzed by
\scite{wang+etal} confirm this result.

An astonishing total of 93 SNRs have been identified in the galaxy
\cite{matonick2}. Remnants 57 and 54 are consistent with the positions
of sources X2 and X3 seen in figure \ref{fig:n5457xcon}. 
\scite{wang+etal} detected X-ray emission from two further
remnants, but they fall outside our JKT image.

\begin{figure}
\centering
\includegraphics[bb=60 70 570 670,angle=270,scale=0.4]{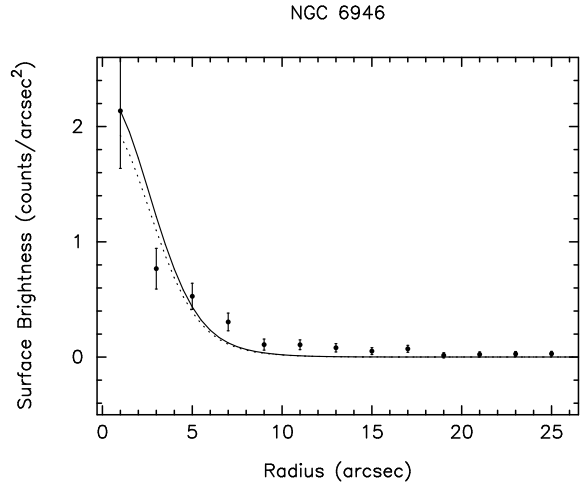}
\includegraphics[bb=60 70 570 670,angle=270,scale=0.4]{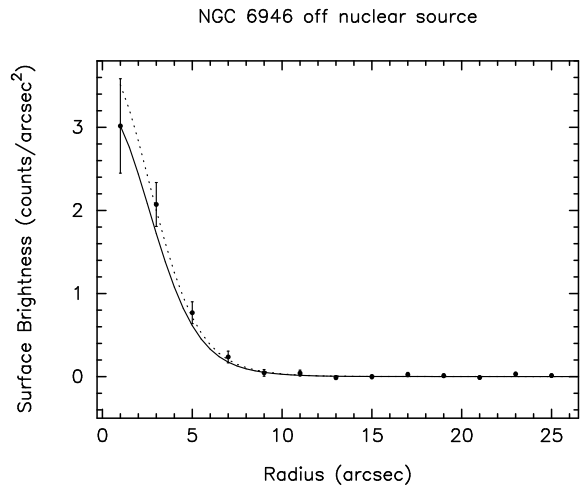}
\caption{Observed profile and model PSF for the nuclear X-3 source
(left) and the off-nuclear X-7 source (right) in NGC\,6946. Models as
in figure \protect\ref{fig:psf.ic342}.}
\label{fig:psf.n6946}
\end{figure}

\begin{figure}
\centering
\setlength{\fboxrule}{1pt}
\setlength{\fboxsep}{0cm}
\fbox{\includegraphics[bb=103 239 508 645,scale=0.5]{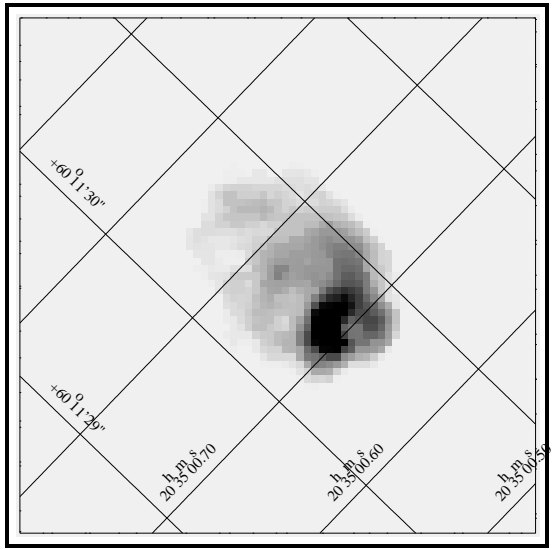}}
\caption{Multiple SNR in NGC\,6946 imagined with the HST PC2.}
\label{fig:n6946_snr}
\end{figure}

{\bf NGC\,6503}: The nucleus of this galaxy has been classified as a
combination of a transition object and a Seyfert 2 nucleus by
\scite{hfs}. The ROSAT HRI observations in figure \ref{fig:n6503xcon}
show extended and elongated emission close to the central region of
the galaxy. This diffuse emission can be better appreciated in figure
\ref{fig:n6503xcon.4} where the X-ray image has been smoothed using a
Gaussians with $\sigma = 8\arcsec$. 

The extended source lies $\sim 10\arcsec$ away from the galactic
nucleus. The astrometric solution of the HRI observation has been
confirmed using two bright X-ray sources in the field with optical
counterparts and therefore the location of the diffuse emission is
probably off-nuclear. Notice, however, that the astrometric checks
carried out in section 3.2.2 assumed a point source distribution of
counts, while the source seen in NGC\,6503 is clearly extended,
implying possible errors in the determination of the centroid of the
emission. The shape and energetics of the diffuse emission could be
explained within the super-wind model for starburst galaxies. More
detailed observations are necessary to confirm this hypothesis.

{\bf NGC\,6946}: Einstein IPC observations showed X-ray emission
associated with the whole body of this galaxy. Two peaks of emission
were detected, one coincident with its starburst nucleus and the other
associated with a prominent northern spiral arm
\cite{fabbiano+trinchieri}. The spectral fit to these data was
consistent with a soft and a hard component, probably associated with
diffuse emission and individual accreting sources respectively. The
presence of diffuse emission across the disk of the galaxy was
confirmed by ROSAT PSPC observations \cite{schlegel2} and can also be
appreciated in our HRI image (figure \ref{fig:n6946xcon}).  The PSPC
also resolved the nucleus into three sources which correspond to X3,
X4 and X7 in the HRI data. Figure \ref{fig:psf.n6946} shows the
profile of the nuclear source X3 and of an off-nuclear source (X7).
Comparing both plots it is clear that the nuclear source is extended
up to $\sim 20\arcsec$ away from the central peak. From the analysis
of ASCA and PSPC data, \scite{ptak+etal2} found that a composite
spectrum (a Raymond-Smith plasma with $kT \sim 1$ keV plus a power-law
with photon index $\Gamma \sim 2.5$) was a good fit to the
observations. It should be remembered, however, that due to the poor
spatial resolution of ASCA most of the sources seen in figure
\ref{fig:n6946xcon} were probably contained in the extraction
apertures of the X-ray spectra.

Six historic SNs have been seen in this spiral galaxy (SNs 1917A,
1939C, 1948B, 1968D, 1969P, and 1980K) \cite{barbon+etal}.  A total of
27 remnants (not including the historic SNs) have been detected by
\scite{matonick2}. SN\,1980K has been observed in X-rays
\cite{schlegel1}, but the source lies outside the $\sim 6\arcmin
\times 6 \arcmin$ JKT optical image of the galaxy. The extreme
northern source seen in this galaxy (X8 in figure \ref{fig:n6946xcon})
seems to belong to an extreme group of SNRs with $\ga 10^{39}$ ergs
s$^{-1}$ \cite{schlegel3}. A faint, red ($R = 18.81; B-V = 1.43$)
counterpart is seen at the position of this source in our optical
images, which also corresponds to the object No 16 in the
\scite{matonick2}'s list of SNRs.  Recent PC2 HST images show that the
optical source has an intricate morphology, with a small bright shell
and what seem to be two outer loops or arcs, as can be seen in the
narrow filter image (F673N) shown in figure \ref{fig:n6946_snr}. Based
on this morphology, \scite{blair} suggested that the source could be
explained as two interacting remnants of different age, with the
smaller shell being a young remnant colliding with the outer, older
shells. However, \scite{chu} have found no high velocity components
in their optical high-dispersion spectra of the source, which is not
consistent with the presence of a very young and compact SNR. From the
H$\alpha$ flux they derived a kinetic energy of $\sim 7 \times
10^{50}$ ergs for the shocked gas. Although this value is somewhat
larger than what is normally assumed for supernova explosions, it is
still within the range of normal events.

\section{Discussion}

First we summarize the most striking results, before discussing some
of the key issues more carefully.

(i) Of the 34 galaxies with X-ray data, 12 have X-ray sources
associated with their nuclear regions: a 35\% success rate. The first
result, then, is that nuclear X-ray sources are very common. (It
should be noticed that due to the astrometric uncertainties discussed
in section 3.2.2, the aligment of the sources with the galactic nuclei
is a tentative result for NGC\,404, IC\,342, and NGC\,4395. However,
the presence of an active nucleus in NGC\,4395 -- see section 5 --
gives further support to this identification).

(ii) However, it is clear that the detection rate changes markedly
with host galaxy luminosity. All 12 detections are in the sub-sample
of 29 galaxies with $M_{B} < -14$. This effect is discussed in more
detail below.

(iii) Of the galaxies spectroscopically classified as Seyfert, LINER,
or transition object, 5/7 have detected nuclear X-ray sources, whereas
only 7/22 objects with HIIR or absorption-line spectra are detected
(spectroscopic classifications will be discussed in Paper III, but the
same result holds using the classifications of \scite{hfs}). However,
this apparent preference for AGN only reflects the fact that nearly
all the smaller galaxies are classified as HIIR. Amongst detected
nuclear X-ray sources, 5/12 are classified as AGN of some kind, and
7/12 as HIIR or absorption line.

(iv) Some of the nuclear sources are only just at the limit of
detectability, but many are considerably more luminous, up to
$10^{40}$ erg s$^{-1}$. Of these more luminous sources, a large
fraction are clearly extended on a scale of 10 arcsec or more,
corresponding to $> 150$ pc or so at the typical distance of our
sample galaxies.

(v) Many sources outside the nucleus are detected -- most
interestingly nine off-nuclear sources are found with luminosities
exceeding $10^{39}$ erg s$^{-1}$, not easily explained as individual
X-ray binaries or SNRs.

\subsection{Dwarf galaxies}

Of the twelve galaxies classified as dwarfs in the sample (see table
\ref{tab:allobs}), only five have Einstein or ROSAT HRI observations
(NGC\,147, NGC\,185, A\,0951+68, Leo\,B and UGC\,6456).  Of these only
one has a positive detection of an X-ray source (UGC\,6456) and even
this one is not nuclear.  For the rest of section 6, we do not include
these dwarfs in the analysis, concentrating on the sub-sample of 29
galaxies with $M_{B} < -14$.

\subsection{Super-luminous X-ray off-nuclear sources}

\begin{table*}
\caption{Luminous off-nuclear  X-ray sources. The numbers in the 
first column correspond to the labels shown in figure
\ref{fig:lx_off_host}.}
\label{tab:offnuc}
\centering
\begin{tabular}{llccl} \hline
Number & Host & ID & $\log L_{X}$ & Remarks \\
& & & (ergs s$^{-1}$) & \\ [10pt]
1 & NGC\,247    &X1    &39.08  &Possible counterparts at the edge\\
  &             &      &       &of a bright star-forming region \\
2 & IC\,342     &X3    &39.18  &Possible ($R \sim 16$) counterpart\\
3 & NGC\,2403   &X1    &39.47  &Possible faint counterpart\\
4 & NGC\,2403   &X3    &39.01  &Possible very faint counterpart - known SNR\\
5 & NGC\,4136   &X1    &39.10  &Diffuse blue counterpart\\
6 & NGC\,5204   &X1    &39.77  &Several faint counterparts\\
7 & NGC\,6946   &X1    &39.13  &No obvious optical counterpart\\
8 & NGC\,6946   &X7    &39.21  &Possible very faint counterparts\\
9 & NGC\,6946   &X8    &39.87  &Known SNR; faint red counterpart\\
\hline
\end{tabular}
\end{table*}

\begin{figure}
\centering
\includegraphics[bb=100 60 560 700,angle=270,scale=0.4]{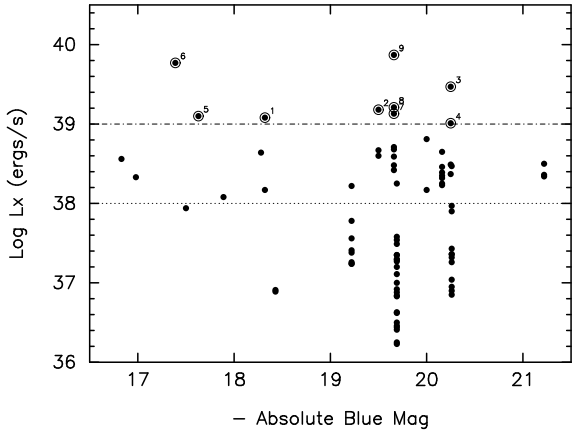}
\caption{X-ray luminosity of off-nuclear sources as a function of the
host absolute blue magnitude. The dotted line across the figure shows
an estimation of the average sensitivity limit for the HRI
observations. Sources with luminosities above $10^{39}$ ergs s$^{-1}$
(dash-dotted line) have been labeled. Comments on their
identifications can be found in table \ref{tab:offnuc}.}
\label{fig:lx_off_host}
\end{figure}

Outside the nucleus, remarkably luminous sources ($L > 10^{38}$ erg
s$^{-1}$) are seen very frequently in galaxies spanning 2 orders of
magnitude in luminosity. These are plotted against host galaxy
luminosity in figure \ref{fig:lx_off_host}. The most luminous sources
($L > 10^{39}$ erg s$^{-1}$) have been labeled with numbers and brief
comments about them can be found in table \ref{tab:offnuc}. Figure
\ref{fig:lx_off_host} shows that there is no obvious correlation with
galaxy size.

\begin{figure}
\centering
\includegraphics[angle=270,scale=0.4]{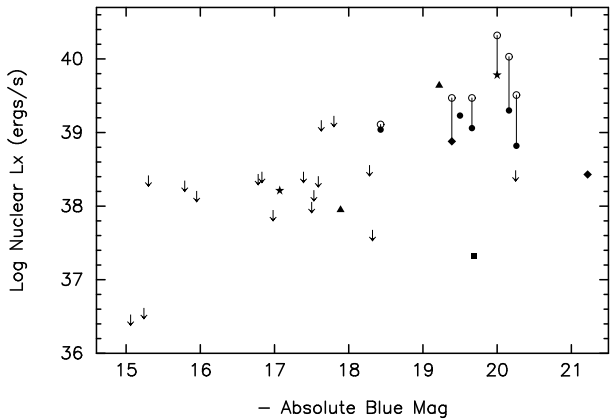}
\caption{Nuclear X-ray luminosity as a function of the host blue
absolute magnitude. Solid symbols show unresolved detections:
starlight dominated nuclei ({\scriptsize \ding{110}}), HIIR nuclei
({\scriptsize \ding{108}}), transition objects (\ding{117}), LINERs
(\ding{72}), and Seyfert nuclei ($\blacktriangle$). Empty circles show
the addition of X-ray flux from an extended component. Arrows
correspond to 2$\sigma$ upper limits.}
\label{fig:lx_host}
\end{figure}

The population of X-ray sources in the Milky Way and M\,31 does not
include luminosities above $10^{38}$ ergs s$^{-1}$. The presence of
luminous ($L_{X} \ga 10^{38}$ ergs s$^{-1}$) sources in the Magellanic
Clouds was assumed to be a metallicity effect
\cite{helfand,vanparadijs+mcclintock}. However, as more galaxies were
surveyed using the capabilities of the Einstein and ROSAT satellites
it became clear that extremely luminous objects ($L_{X} \ga 10^{39}$
ergs s$^{-1}$) were not rare \cite{fabbiano2}. 

Next we assess the possibility that the ultra-luminous sources are
actually background objects. Using the results from the ROSAT Deep
Survey in the Lockman Field \cite{hasinger} we estimate that no more
than three background sources with fluxes of $10^{-12} - 10^{-13}$
ergs s$^{-1}$ cm$^{-2}$ would be found in a 1 deg$^{2}$ field. If
these targets are distant, and so obscured by the intervening galaxy,
their intrinsic luminosities must be higher and the corresponding
number densities even lower.  Given the average projected size of the
galaxies on the sky (see table \ref{tab:nh_conv} for the area
contained within the D$_{25}$ ellipses) the probability that all the
sources correspond to background objects is extremely small.

At least one super-luminous source has been identified as a multiple
object formed by interacting SNRs in the disk of NGC\,6946
\cite{blair} (see figure \ref{fig:n6946_snr}). \scite{wang} has 
recently suggested that some of the X-ray bright remnants observed in
M\,101 could correspond to {\it `Hypernova Remnants'\/}, a term
introduced by Paczy\'{n}ski (1998) for super-energetic $\gamma$-ray
bursts and the associated afterglow event - but see also \scite{chu}.
Other multiple object systems could be formed from XRBs, SNRs, and
diffuse emission from hot bubbles of interstellar gas. Alternatively,
the sources could be super-Eddington XRBs with luminosities several
times greater than the Eddington limit for a $\sim 1.4 M_{\odot}$
neutron star.  The high luminosities in this case are explained as
anisotropic emission from neutron stars in binary systems with very
strong magnetic fields \cite{vanparadijs+mcclintock}, or as black hole
XRBs.

\begin{figure}
\centering
\includegraphics[bb=150 110 560 640,angle=270,scale=0.4]{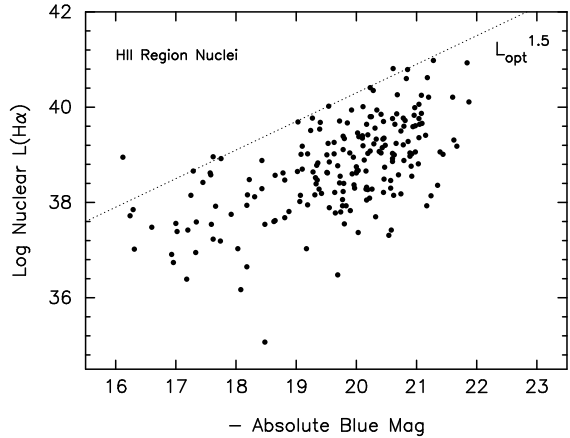}
\caption{Nuclear H$\alpha$ luminosities versus host galaxy absolute
magnitude for galaxies with HII region nuclei. Data from \protect\scite{hfs2}.}
\label{fig:hiir_la_host}
\end{figure}

\begin{figure}
\centering
\includegraphics[bb=150 110 560 640,angle=270,scale=0.4]{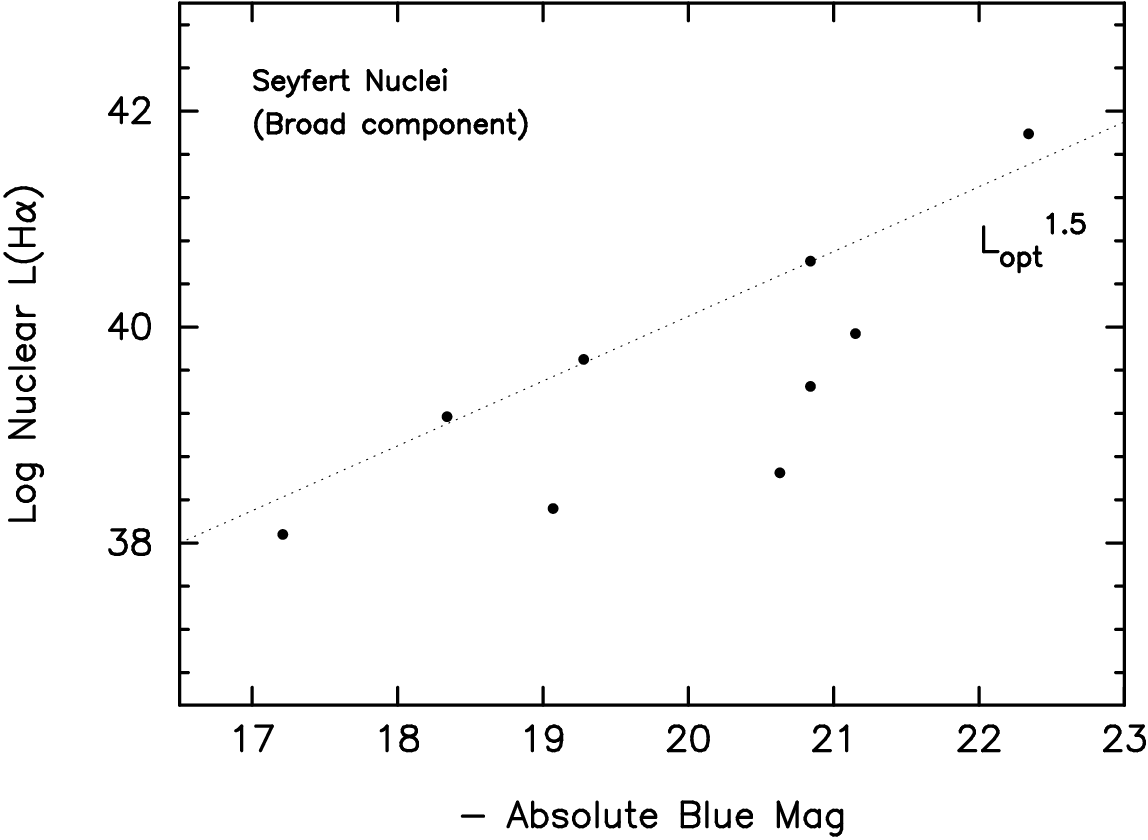}
\caption{Nuclear broad H$\alpha$ luminosities versus host galaxy absolute 
magnitude for galaxies with Seyfert nuclei. Data from \protect\scite{hfs2}.}
\label{fig:seyfert.b_la_host}
\end{figure}

\begin{figure}
\centering
\includegraphics[bb=150 110 560 640,angle=270,scale=0.4]{sey.n_la_host.ps}
\caption{Nuclear narrow H$\alpha$ luminosities versus host galaxy absolute 
magnitude for galaxies with Seyfert nuclei. Data from \protect\scite{hfs2}.}
\label{fig:seyfert_la_host}
\end{figure}

\begin{figure}
\centering
\includegraphics[bb=150 110 560 640,angle=270,scale=0.4]{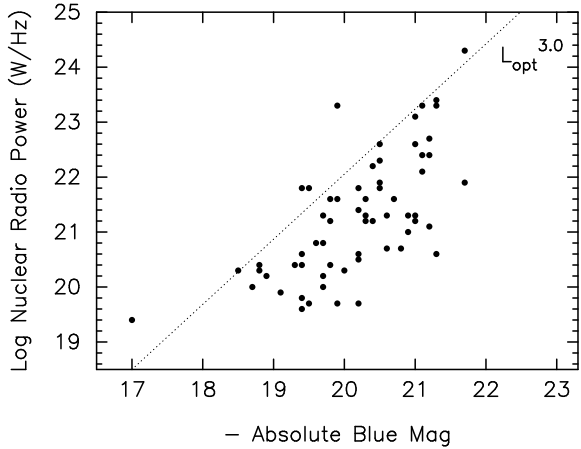}
\caption{Nuclear radio power as a function of host galaxy absolute magnitude 
(adapted from \protect\scite{sadler}).}
\label{fig:lr_host}
\end{figure}

\scite{colbert} have recently analyzed ASCA data from three nearby
spiral galaxies. One of the galaxies is M\,33 and these observations
have already been discussed in section 5. The other two galaxies
(NGC\,1313 and NGC\,5408) each harbor a luminous source ($L_{X} \ga
10^{39}$ ergs s$^{-1}$) displaced $\sim 50 \arcsec$ from the
nucleus. The fit to the X-ray spectra of these sources shows that at
least two components, a steep power law and a `Disk Black Body', are
required to explain the observations, suggesting that these are
accretion driven systems. The parameters from the model imply BH
masses of $100 - 10,000 M_{\sun}$, intermediate between those seen in
XRB and AGN. Given the observed X-ray luminosities and the estimated
masses, these objects are not super Eddington sources.

It is clear, then, that super-luminous point sources are common.
However, unlike nuclear sources discussed in the next section, there
seems to be no correlation between the probability of having a
super-luminous off-nuclear source and the brightness of the parent
galaxy.

\subsection{Correlation between nuclear X-ray luminosity and host 
galaxy luminosity}

Figure \ref{fig:lx_host} shows the nuclear X-ray luminosities and
upper limits for the galaxies in the sample as a function of the blue
absolute magnitude of the host. Different symbols have been used to
show the optical classification of the nuclear emission as starlight,
HII regions, LINERs, transition objects (i.e., a HIIR and LINER
composite), and Seyferts (Paper III). As was suggested before, it is
found that the probability of detection of a nuclear X-ray source
correlates strongly with host galaxy luminosity: the rate of detection
is 0\% (0/7) below $M_{B} \sim -17$, 25\% (3/12) for $-17 < M_{B} <
-19$, and 90\% (9/10) for galaxies with $M_{B} < -19$.

\scite{colbert} have collected data from ROSAT HRI observations for 39 
nearby spiral and elliptical galaxies, of which 9 objects are common
to our sample. Their results confirm what is seen in figure
\ref{fig:lx_host}: for those sources detected within $\sim 6$ arcsecs
from the optical center of the galaxy, and therefore likely to be
nuclear sources, the detection rate is nearly 100 percent for hosts
with $M_{B} \la -20$.

The distribution of points does not suggest a straightforward
correlation, but rather a large spread with an upper envelope. In
other words, large galaxies can have luminous or feeble nuclei, but
small galaxies can only have feeble nuclei. A similar upper-envelope
effect has been claimed for the host galaxy luminosities of quasars
and Seyferts \cite{mcleod3}. Given this correlation, the lack of
detection of nuclear sources in small galaxies does not necessarily
mean they cannot form AGN at all -- it may be that they are
exceedingly weak nuclear sources.

It is not clear how to model the envelope-like relationship between
nuclear and host galaxy luminosities. However, a simple visual
comparison with other related samples reveals a rather striking
effect. The upper envelope for our sample seems to go roughly as
$L_{x} \propto L_{\rm host}^{1.5}$. Figures \ref{fig:hiir_la_host},
\ref{fig:seyfert.b_la_host} and \ref{fig:seyfert_la_host} show nuclear
emission line data taken from the study of \scite{hfs}. The nuclear
H$\alpha$ luminosity from galaxies classified as HII regions shows a
very similar correlation with host galaxy luminosity (figure
\ref{fig:hiir_la_host}) and once again an upper envelope with slope
1.5 is a good fit. Likewise for dwarf Seyferts, the broad H$\alpha$
component shows a correlation with a slope of 1.5 (figure
\ref{fig:seyfert.b_la_host}). However, for the {\em narrow-line}
component of H$\alpha$ for Seyfert nuclei, the correlation is much
steeper, with a slope of 3.0 (figure
\ref{fig:seyfert_la_host}). Figure \ref{fig:lr_host} shows data from
the radio survey of elliptical galaxies by \scite{sadler}. Here we
find that nuclear radio luminosity shows a very strong correlation
with an upper envelope slope of 3.0. (\scite{sadler} do a more careful
analysis showing that the 30th percentile goes as $L_{\rm
host}^{2.2}$). So it seems that narrow emission lines in AGN are
intimately connected with radio emission. On the other hand, the
relation seen for our X-ray sources is equally consistent with either
AGN or star formation.

\subsection{Nuclear X-ray sources as disk sources}

Some of the nuclear X-ray sources we have seen are consistent with the
most luminous known X-ray binaries. Furthermore, we have seen that
super-luminous off-nuclear sources do occur quite often. Is it
possible, then, that the nuclear sources are simply examples of the
general X-ray source population that happen to be located in the
nucleus?  Where we see a clear extended source, this cannot be the
case. This argument assumes that the observed extended emission is
truly diffuse in nature. Of course, we cannot exclude a contribution
of unresolved sources to the extended component, but this cannot
account for the bulk of the emission for the more conspicuous cases
(see \pcite{dahlem+etal}, \pcite{roberts+etal}, \pcite{ehle}, and
\pcite{schlegel2}). It could also be that more luminous galaxies are
simply more likely to have at least one example of a particularly
luminous source. The strongest argument against this is that amongst
the off-nuclear sources we do not see the upper-envelope correlation
with host galaxy luminosity (figure \ref{fig:lx_off_host}). We can
also crudely estimate the likely size of such an effect if we know the
luminosity distribution of X-ray sources.

The {\it combined\/} luminosity distribution of disk (ie, off-nuclear)
X-ray sources from a sample of 83 spiral galaxies has been established
recently by \scite{roberts+warwick}. The sample was defined as those
objects surveyed by \scite{hfs} that had archive ROSAT HRI
observations. The distribution determined by \scite{roberts+warwick}
reaches luminosities for off-nuclear sources just below $10^{41}$ ergs
s$^{-1}$, ie, it extends well into the realm of the super-luminous
sources discussed in the previous section. The low luminosity end of
the distribution was found using the ROSAT PSPC deep observations of
NGC\,224 (M\,31) published by \scite{supper}. \scite{roberts+warwick}
find that the (differential) luminosity distribution is well fitted by
a power law slope of $-1.8$, with a flattening for luminosities $\la
10^{36}$ ergs s$^{-1}$. It is not clear whether this change in the
slope is due to incompleteness or to an intrinsic variation in the
faint source population.

Since the luminosity distribution determined by
\scite{roberts+warwick} corresponds to an average distribution a
direct comparison between the properties of the populations in
individual galaxies and their hosts is not possible. Moreover, as
their sample is drawn from the flux-limited sample surveyed by
\scite{hfs}, with an under-representation of low luminosity galaxies,
it is not possible to confirm the lack of correlation between the
probability of finding a super-luminous off-nuclear source and the
luminosity parent galaxy seen in figure \ref{fig:lx_off_host}.

The luminosity distribution of off-nuclear sources predicts that about
8 disk sources with $L_{X} \geq 10^{37}$ ergs s$^{-1}$ will be found
in a $10^{10} \times L_{\sun}$ ($M \sim -20$) galaxy. Of these sources
only one will have a luminosity $\geq 10^{38}$ ergs s$^{-1}$. The
probability of that source being located within the nucleus of the
parent galaxy can be expressed as the ratio between the total
luminosity and the luminosity of a `nuclear region' of 100 pc (equal
to the HRI spatial resolution at a distance of 4 Mpc). Using the well
established exponential disk profile in spirals, and assuming a scale
length for the disk equal to 4 kpc \cite{simien} this ratio is only
$\sim 10^{-4}$.

For more luminous sources the situation is even worse, since the
\scite{roberts+warwick} luminosity distribution predicts that about $8
\times 10^{10} \times L_{\sun}$ galaxies have to be surveyed in order
to find a $L_{X} \geq 10^{39}$ ergs s$^{-1}$ X-ray source. This
prediction does not agree with our results, however. In our sample of
29 galaxies we have detected 9 sources with luminosities above
$10^{39}$ ergs s$^{-1}$ and all but one of the host galaxies
(NGC\,2403) have $M_{B} > -20$ (see figure \ref{fig:lx_off_host}). A
closer look at the results reveals that two effects could be
responsible for this difference: (1) statistic fluctuations introduced
in the results drawn from our smaller sample due to the distances
adopted for the parent galaxies; (2) the fairly large threshold radius
(25 arcsecs) adopted by \scite{roberts+warwick} to discriminate
between nuclear and off-nuclear sources (as an example, the
ultra-luminous off-nuclear source in NGC\,5204 was labeled as nuclear
by \pcite{roberts+warwick}). As pointed by \scite{roberts+warwick},
individual galaxies can also show important deviations from the
determined luminosity distribution. The X-ray population in NGC\,5457,
for example, shows the same power-law index determined by
\scite{roberts+warwick} for the combined sample, but the normalization
is 4 times larger.

In summary, even with a large error (or fluctuation) in the
normalization of the luminosity distribution of off-nuclear sources,
it is clear that nuclear sources cannot be explained as disk sources
located {\it by chance\/} in the nuclei of galaxies. There is another
possibility, however. Nuclear sources could be explained if massive
X-ray emitting systems, such as black hole binaries, suffer
considerable dynamical friction against field stars and so migrate to
the centre of their parent galaxies. In this scenario the upper
envelope seen in figure \ref{fig:lx_host} could be the result of
smaller galaxies having shallower potential wells and therefore being
less efficient in dragging the heavy objects to their
centres. However, it is well known that the scale time for a
significant change in the orbit of any stellar system is comparable to
the local relaxation time which, in the case of disks, is much longer
than the Hubble time \cite{binney+tremaine}. Therefore, galactic disks
are too young for a source to have migrated significantly from its
original location.
 
\subsection{Nuclear X-ray sources as bulge sources}

An alternative to the scenario described above is that the massive
XRBs could have originally been located in the bulge of the galaxy,
where larger background stellar densities make mass segregation a much
more efficient mechanism. However, the anticorrelation between the
time it takes for a massive object to migrate to the galactic nucleus
and the mass of the object implies that an XRB would need a mass of
over $100 M_{\sun}$ to fall to the galactic centre from a distance of
only 10 pc within the lifetime of the parent galaxy (1 Gyr)
\cite{morris}. Therefore, dynamical friction will only be efficient
for very heavy or very nearby stars. Indeed, \scite{binney+tremaine}
show that only extremely massive objects such as $10^{6} M_{\sun}$
globular clusters would have spiralled into the centre of their parent
galaxies in less than a Hubble time from a significantly extra-nuclear
(2 kpc) distance.

It could also be argued that the X-ray emission from the central
regions corresponds to bulge sources which appear coincident with the
galactic nuclei. This is unlikely, however, given the spatial
resolution of our observations (ranging from $\sim 16$ pc for NGC\,224
to $\sim 190$ pc for NGC\,5457, with most galaxies with detected
nuclear sources being observed at a resolution $\ga 100$ pc), implying
that a very large number of highly centrally concentrated bulge
sources would be necessary to explain the observations.  NGC\,224
indeed shows a high concentration of bulge sources when compared with
our own Galaxy, with 4 times more sources seen in NGC\,224 within the
central 5 arcmins \cite{primini}. However, only two sources are found
within the innermost 100 pc, giving a total luminosity of $4 \times
10^{37}$ ergs s$^{-1}$. The probability of those sources having $L_{X}
\ga 10^{38}$ ergs s$^{-1}$ is extremely small, as is shown in the
bulge luminosity distribution determined of \scite{primini}.  Also, as
we have already shown, the hypothesis that particularly heavy (and
luminous) binary systems are located within this region because of
mass segregation is not viable.  Therefore, most of the galaxies with
$M_{B} < -18$ seen in figure \ref{fig:lx_host} would need to harbor a
much larger population of bulge sources that the one seen in NGC\,224
in order to explain their observed nuclear luminosities, which is a
very unlikely scenario.

\subsection{The nature of the nuclear X-ray sources}

The very different correlations between X-ray luminosity and host
galaxy absolute magnitude seen in figures \ref{fig:lx_off_host} and
\ref{fig:lx_host} for nuclear and off-nuclear sources imply 
two different populations. Nuclear sources are not disk or bulge
sources located in the nuclear region by chance. Instead, nuclear
sources have a particular {\it nuclear\/} nature.

The next step is to try to unveil this nature. What are these sources?
Are they connected with stellar processes in the nuclear region of the
galaxies? Could they be an expression of nuclear activity that somehow
escapes detection at optical wavelengths?

An interesting pattern can be seen in figure \ref{fig:lx_host} between
the optical (spectral) classification of the galaxies and their X-ray
luminosities. Both LINER objects (NGC\,404 and NGC\,4258) lie close to
the upper envelope of the distribution of detected sources, as does
the Seyfert galaxy NGC\,3031. The second Seyfert nucleus, NGC\,4395,
is heavily absorbed below 3 keV and its intrinsic soft X-ray
luminosity is estimated to be an order of magnitude larger than the
values found from ROSAT observations \cite{iwasawa}. Introducing this
last correction all objects classifies as AGN would be found near the
top of the upper envelope in figure \ref{fig:lx_host}.

\begin{figure}
\centering
\includegraphics[bb=150 110 560 640,angle=270,scale=0.4]{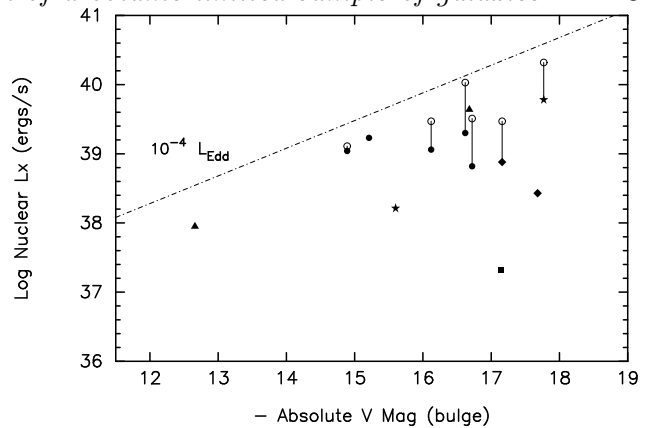}
\caption{Nuclear X-ray luminosity as a function of the host visual
absolute magnitude of the bulge component. Symbols as in figure
\ref{fig:lx_host}. The dash-dotted line represents $10^{-4}$ times the predicted 
X-ray emission from central black hole masses at the Eddington limit
assuming the bulge to BH relationship determined by
\protect\scite{vandermarel1} and an X-ray to bolometric flux ratio
of 0.1 (see text).}
\label{fig:lx_blg_host}
\end{figure}

The distribution of X-ray sources with galaxy absolute magnitude,
shown in figure \ref{fig:lx_host}, can be bounded by a nearly linear
upper envelope. A similar linear upper envelope has been found in the
correlation between nuclear (AGN) luminosity and host galaxy
luminosity for samples of Seyfert galaxies and low redshift quasars at
IR and optical wavelengths \cite{yee.92,mcleod}. This suggests that
there is a maximum allowed AGN luminosity which is an increasing
function of the luminosity of the parent galaxy. The limit could be
the result of a correlation between the total mass of the host galaxy
and the mass of the central engine. The observed range of nuclear
luminosities would be given by different accretion rates, with sources
located at the bound of the region, showing the maximum possible
luminosity. Indeed, from their study of nearby quasars \scite{mcleod2}
and \scite{mclure} show that if the correlation between black hole
mass and galaxy size claimed by \scite{magorrian+etal} is correct,
then the upper envelope is consistent with the most luminous AGN
radiating at roughly 20\% of the Eddington limit. If the smaller ratio
of black hole mass to galaxy mass suggested by \scite{vandermarel1} is
used then the closeness of the upper envelope to the Eddington
prediction is even more remarkable.

However, in the case of the comparatively weak nuclear X-ray sources
we have found here, it is not clear what process determines the
maximum possible luminosity. If the observed X-ray emission is driven
by accretion of matter onto a BH, then the Eddington limit for the
central mass seems the most straight-forward mechanism to explain the
existence of the linear upper envelope. However, this cannot be the
case. The only two Seyfert galaxies in the sample, NGC\,3031 (M\,81)
and NGC\,4395, are thought to be extremely sub-Eddington
\cite{hfs.96,lira}, with their bolometric luminosities being only
$10^{-4} - 10^{-3} \times L_{\rm Edd}$.  

Figure \ref{fig:lx_blg_host} makes this point graphically. We use the
relation given by \scite{vandermarel1}, $\log M_{BH} = -1.96 + \log
L$, where $L$ is the $V$-band luminosity of the bulge and all
quantities are expressed in solar units. The bulge magnitudes for the
galaxies in our sample can be found from the total magnitudes shown in
table \ref{tab:allobs} using the relantionship obtained by
\scite{simien} as a function of Hubble Type, while a colour $B-V =1.0$
was assumed for all objects to convert the blue magnitudes to visual
magnitudes. Using the relationship above we can infer the X-ray
output for a central BH emitting at the Eddington limit as a function
of its host galaxy bulge luminosity, assuming that the soft X-ray
emission corresponds to $\sim 10\%$ of the bolometric luminosity. It
is found that the predicted X-ray luminosities (seen as a dotted line
in figure \ref{fig:lx_blg_host}) are about 4 orders of magnitude
brighter than what is observed.

The nuclear-host correlation we have seen cannot, therefore, be simply
interpreted as the upper envelope relation seen amongst luminous AGN,
because the X-ray sources are so feeble. However, it should be borne
in mind that the objects discussed in \scite{mcleod2} have been
pre-selected as AGN, whereas we are looking at galaxies as a
whole. For example, it is possible that the probability distribution
$P(L_{x}/L_{\rm Edd})$ is a universal function for all galaxies,
declining continuously from large values at small $L_{x}/L_{\rm Edd}$
to small values at large $L_{x}/L_{\rm Edd}$, but with a cutoff at
$L_{x}/L_{\rm Edd} = 1$. The envelope seen in AGN samples then traces
the cutoff, but the envelope seen in complete galaxy samples is a
statistical effect, tracing in each host-galaxy-bin the nuclear
luminosity beyond which the expected number of objects is just less
than one. With a larger sample, the density of points would be
greater, so the envelope would move up, but the slope might stay the
same.

Most of the preceding discussion has centred round the possibility
that the weak nuclear X-ray sources are AGN.  However, this is not
obvious. They seem to show no particular preference for AGN-like
optical spectra, and the correlation with host galaxy luminosity is
exactly like that shown by the emission line luminosity for HII nuclei
(see figure \ref{fig:hiir_la_host}). Furthermore many of the most
luminous sources are extended on scales of hundreds of parsecs or
more. It seems equally likely that the X-ray emission is connected
with star formation, but with the current data we cannot go much
further. In later papers in this series we examine the correlation
between X-ray, optical, and emission-line luminosities, and test
specific AGN and star formation models.

Although it is unclear which physical process causes the observed
distributions, the upper envelopes give strong evidence for a
correlation between the host galaxy luminosity and the level of
nuclear activity. The fact that an upper envelope is also observed in
the distribution of the nuclear HII regions might suggest that the
correlations are governed by the amount of gas available in the
nuclear regions. If bigger galaxies are more efficient in dragging gas
to their nuclear regions they could show more vigorous star formation
and, potentially, feed their massive BHs more efficiently.

\section{Summary}

X-ray sources in the nuclei of galaxies are very common and the
luminosity of these sources is strongly connected to the luminosity of
the host galaxy.  The highest luminosities reach $\sim 10^{40}$ ergs
s$^{-1}$ which is $10^{4}$ times less than the Eddington limit for the
massive black holes that may be present in such galaxies. The most
luminous nuclear X-ray sources are frequently extended on scales of
hundreds of parsecs. It is, so far, unclear whether these nuclear
X-ray sources are miniature AGN of some kind or a phenomenon connected
with normal star formation.  Outside the nuclei of galaxies, extremely
luminous sources ($L > 10^{39}$ erg s$^{-1}$, compared with `normal'
$\sim 10^{37}$ erg s$^{-1}$ SNRs or $\sim 10^{38}$ erg s$^{-1}$ XRBs)
are also quite common, occurring in a third of all galaxies. The
nature of these sources is currently also unclear.

\section*{Acknowledgements}

We thank Gordon Stewart and John Arabadjis for insightfull discussion
on this paper.  This research has made use of data obtained from the
Leicester Database and Archive Service at the Department of Physics
and Astronomy, Leicester University, UK. The Jacobus Kapteyn Telescope
is operated on the island of La Palma on behalf of the UK Particle
Physics and Astronomy Research Council (PPARC) and the Nederlanse
Organisatie voor Wetenschappelijk Onderzoek (NWO) and is the located
at the Spanish Observatorio del Roque de los Muchachos. The
observatory is hosted by the Instituto de Astrof\'{\i}sica de
Canarias. The DSS plates used in this paper were based on photographic
data of the National Geographic Society -- Palomar Observatory Sky
Survey (NGS-POSS) obtained using the Oschin Telescope on Palomar
Mountain.  The NGS-POSS was funded by a grant from the National
Geographic Society to the California Institute of Technology.  The
plates were processed into the present compressed digital form with
their permission.  The Digitized Sky Survey was produced at the Space
Telescope Science Institute under US Government grant NAG W-2166.  The
data presented in this paper were reduced using Starlink facilities.

\bibliographystyle{my_mnras}
\bibliography{paper_figs}

\begin{thebibliography}{{McLeod}, {Rieke} \& {Storrie-Lombardi}<1999>}

\bibitem[{Barbon}, {Capellaro} \& {Turatto}<1989>]{barbon+etal}
{Barbon} R., {Capellaro} E., {Turatto} M., 1989,
\newblock A\&AS, {81}, 421

\bibitem[{Bender}, {Kormendy} \& {Dehnen}<1996>]{bender}
{Bender} R., {Kormendy} J., {Dehnen} W., 1996,
\newblock ApJ, {464}, L123

\bibitem[{Binney} \& {Tremaine}<1987>]{binney+tremaine}
{Binney} J., {Tremaine} S., 1987,
\newblock Galactic dynamics

\bibitem[{Blair}, {Fesen} \& {Schlegel}<1997>]{blair}
{Blair} W.~P., {Fesen} R.~A., {Schlegel} E.~M., 1997,
\newblock Bull. American Astron. Soc., {190}, 2706

\bibitem[{Bohlin} {et~al.}<1983>]{bohlin}
{Bohlin} R.~C., {Cornett} R.~H., {Hill} J.~K., {Smith} A.~M., {Stecher} T.~P.,
  1983,
\newblock ApJ, {274}, L53

\bibitem[{Boller} {et~al.}<1992>]{boller}
{Boller} T., {Meurs} E. J.~A., {Brinkmann} W., {Fink} H., {Zimmermann} U.,
  {Adorf} H.~M., 1992,
\newblock A\&A, {261}, 57

\bibitem[{Brandt} {et~al.}<1997>]{brandt}
{Brandt} W.~N., {Ward} M.~J., {Fabian} A.~C., {Hodge} P.~W., 1997,
\newblock MNRAS, {291}, 709

\bibitem[{Bregman}, {Cox} \& {Tomisaka}<1993>]{bregman}
{Bregman} J.~N., {Cox} C.~V., {Tomisaka} K., 1993,
\newblock ApJ, {415}, L79

\bibitem[{Chu}, {Chen} \& Shih-Ping<1999>]{chu}
{Chu} Y.-H., {Chen} C.-H.~R., Shih-Ping L., 1999,
\newblock in STScI Symposium: The Largest Explosions Since the Big Bang:
  Supernovae and Gamma Ray Bursts,
\newblock preprint, astro-ph/9909091

\bibitem[{Colbert} \& {Mushotzky}<1999>]{colbert}
{Colbert} E. J.~M., {Mushotzky} R.~F.,
\newblock 1999,
\newblock preprint, astro-ph/9901023

\bibitem[{Dahlem}, {Weaver} \& {Heckman}<1998>]{dahlem+etal}
{Dahlem} M., {Weaver} K.~A., {Heckman} T.~M., 1998,
\newblock apjs, {118}, 401

\bibitem[{David} {et~al.}<1998>]{david+etal}
{David} L.~P., {Harnden} F.~R., {Kearns} K.~E., {Zombeck} M.~V., 1998,
\newblock The ROSAT High Resolution Imager (HRI) Calibration Report

\bibitem[{Dibai} \& {Zasov}<1985>]{dibai+zasov}
{Dibai} E.~A., {Zasov} A.~V., 1985,
\newblock Soviet Astronomy, {29}, 273

\bibitem[{Drissen} \& {Roy}<1996>]{drissen+roy}
{Drissen} L., {Roy} J.~R., 1996,
\newblock in ASP Conf. Ser. 98: From Stars to Galaxies: the Impact of Stellar
  Physics on Galaxy Evolution, eds, {Leitherer} C., {Fritze-Von Alvensleben}
  U., {Huchra} J.

\bibitem[{Ehle} {et~al.}<1998>]{ehle}
{Ehle} M., {Pietsch} W., {Beck} R., {Klein} U., 1998,
\newblock A\&A, {329}, 39

\bibitem[{Fabbiano} \& {Panagia}<1983>]{fabbiano+panagia}
{Fabbiano} G., {Panagia} N., 1983,
\newblock ApJ, {266}, 568

\bibitem[{Fabbiano} \& {Trinchieri}<1984>]{fabbiano+trinchieri2}
{Fabbiano} G., {Trinchieri} G., 1984,
\newblock ApJ, {286}, 491

\bibitem[{Fabbiano} \& {Trinchieri}<1987>]{fabbiano+trinchieri}
{Fabbiano} G., {Trinchieri} G., 1987,
\newblock ApJ, {315}, 46

\bibitem[{Fabbiano}, {Kim} \& {Trinchieri}<1992>]{fabbiano+etal}
{Fabbiano} G., {Kim} D.~W., {Trinchieri} G., 1992,
\newblock ApJS, {80}, 531

\bibitem[{Fabbiano}<1988>]{fabbiano}
{Fabbiano} G., 1988,
\newblock ApJ, {325}, 544

\bibitem[{Fabbiano}<1995>]{fabbiano2}
{Fabbiano} G., 1995,
\newblock in X-ray binaries, eds, {Lewin} W.~H., {Van Paradijs} J., {Van Den
  Heuvel} E.~P.

\bibitem[{Filippenko} \& {Sargent}<1989>]{filippenko+sargent}
{Filippenko} A.~V., {Sargent} W. L.~W., 1989,
\newblock ApJ, {342}, L11

\bibitem[{Gallais} {et~al.}<1991>]{gallais}
{Gallais} P., {Rouan} D., {Lacombe} F., {Tiphene} D., {Vauglin} I., 1991,
\newblock A\&A, {243}, 309

\bibitem[{Gehren} {et~al.}<1984>]{gehren+etal}
{Gehren} T., {Fried} J., {Wehinger} P.~A., {Wyckoff} S., 1984,
\newblock ApJ, {278}, 11

\bibitem[{Hasinger} {et~al.}<1998>]{hasinger}
{Hasinger} G., {Burg} R., {Giacconi} R., {Schmidt} M., {Trumper} J., {Zamorani}
  G., 1998,
\newblock A\&A, {329}, 482

\bibitem[{Helfand}<1984>]{helfand}
{Helfand} D.~J., 1984,
\newblock PASP, {96}, 913

\bibitem[{Ho}, {Filippenko} \& {Sargent}<1995>]{hfs}
{Ho} L.~C., {Filippenko} A.~V., {Sargent} W. L.~W., 1995,
\newblock ApJS, {98}, 477

\bibitem[{Ho}, {Filippenko} \& {Sargent}<1996>]{hfs.96}
{Ho} L.~C., {Filippenko} A.~V., {Sargent} W. L.~W., 1996,
\newblock ApJ, {462}, 183

\bibitem[{Ho}, {Filippenko} \& {Sargent}<1997a>]{hfs2}
{Ho} L.~C., {Filippenko} A.~V., {Sargent} W. L.~W., 1997a,
\newblock ApJS, {112}, 315

\bibitem[{Ho}, {Filippenko} \& {Sargent}<1997b>]{hfs3}
{Ho} L.~C., {Filippenko} A.~V., {Sargent} W. L.~W., 1997b,
\newblock ApJ, {487}, 568

\bibitem[{Huchra} \& {Burg}<1992>]{huchra+burg}
{Huchra} J., {Burg} R., 1992,
\newblock ApJ, {393}, 90

\bibitem[{Hutchings}, {Crampton} \& {Campbell}<1984>]{hutchings+etal}
{Hutchings} J.~B., {Crampton} D., {Campbell} B., 1984,
\newblock ApJ, {280}, 41

\bibitem[{Ishisaki} {et~al.}<1996>]{ishisaki}
{Ishisaki} Y., {Makishima} K., {Iyomoto} N., {Hayashida} K., {Inoue} H.,
  {Mitsuda} K., {Tanaka} Y., {Uno} S.~I., {Kohmura} Y., {Mushotzky} R.~F.,
  {Petre} R., {Serlemitsos} P.~J., {Terashima} Y., 1996,
\newblock PASJ, {48}, 237

\bibitem[{Iwasawa} {et~al.}<2000>]{iwasawa}
{Iwasawa} K., {Fabian} A.~C., {Almaini} O., {Lira} P., {Lawrence} A.,
  {Hayashida} K., {Inohue} H.,
\newblock 2000,
\newblock submitted

\bibitem[Johnson<1997>]{johnson}
Johnson R.~A., 1997,
\newblock {\it PhD thesis}, Queen Mary and Westfiled College

\bibitem[{Kim}, {Fabbiano} \& {Trinchieri}<1992>]{kim+fabbiano+trinchieri}
{Kim} D.~W., {Fabbiano} G., {Trinchieri} G., 1992,
\newblock ApJ, {393}, 134

\bibitem[{Kormendy} \& {McClure}<1993>]{kormendy+mcclure}
{Kormendy} J., {McClure} R.~D., 1993,
\newblock AJ, {105}, 1793

\bibitem[{Kormendy} \& {Richstone}<1995>]{kormendy+richston}
{Kormendy} J., {Richstone} D., 1995,
\newblock ARA\&A, {33}, 581

\bibitem[{Kraan-Korteweg} \& {Tammann}<1979>]{kraan-korteweg+tamman}
{Kraan-Korteweg} R.~C., {Tammann} G.~A., 1979,
\newblock Astron. Nachr., {300}, 181

\bibitem[{Kraan-Korteweg}<1986>]{kraan-korteweg}
{Kraan-Korteweg} R.~C., 1986,
\newblock A\&AS, {66}, 255

\bibitem[{Kraft}, {Burrows} \& {Nousek}<1991>]{kraft}
{Kraft} R.~P., {Burrows} D.~N., {Nousek} J.~A., 1991,
\newblock ApJ, {374}, 344

\bibitem[{Kunth}, {Sargent} \& {Bothun}<1987>]{kunth+etal}
{Kunth} D., {Sargent} W. L.~W., {Bothun} G.~D., 1987,
\newblock AJ, {93}, 29

\bibitem[{Lira} {et~al.}<1999>]{lira}
{Lira} P., {Lawrence} A., {O'Brien} P., {Johnson} R.~A., {Terlevich} R.,
  {Bannister} N., 1999,
\newblock MNRAS, {305}, 109

\bibitem[{Loewenstein} {et~al.}<1998>]{loewenstein}
{Loewenstein} M., {Hayashida} K., {Toneri} T., {Davis} D.~S., 1998,
\newblock ApJ, {497}, 681

\bibitem[{Long} {et~al.}<1994>]{long+etal}
{Long} K.~S., {Gordon} S.~M., {Blair} W.~P., {Charle} P.~A., 1994,
\newblock in AIP Conference Proceedings: The soft X-ray cosmos, eds, {Schlegel}
  E.~M., {Petre} R.

\bibitem[{Lynds} {et~al.}<1998>]{lynds}
{Lynds} R., {Tolstoy} E., {O'Neil.}, Earl~J. J., {Hunter} D.~A., 1998,
\newblock AJ, {116}, 146

\bibitem[{Mackie} {et~al.}<1995>]{mackie+etal}
{Mackie} G., {Fabbiano} G., {Kim} D.-W., {Ikebe} Y., 1995,
\newblock in IAU Sym. 164, Stellar Populations, eds, {van der Kruit} P.~C.,
  {Gilmore} G.

\bibitem[{Magorrian} {et~al.}<1998>]{magorrian+etal}
{Magorrian} J., {Tremaine} S., {Richstone} D., {Bender} R., {Bower} G.,
  {Dressler} A., {Faber} S.~M., {Gebhardt} K., {Green} R., {Grillmair} C.,
  {Kormendy} J., {Lauer} T., 1998,
\newblock AJ, {115}, 2285

\bibitem[{Makishima} {et~al.}<1989>]{makishima}
{Makishima} K., {Ohashi} T., {Hayashida} K., {Inoue} H., {Koyama} K., {Takano}
  S., {Tanaka} Y., {Yoshida} A., {Turner} M. J.~L., {Thomas} H.~D., {Stewart}
  G.~C., {Williams} R.~O., {Awaki} H., {Tawara} Y., 1989,
\newblock PASJ, {41}, 697

\bibitem[{Maoz} {et~al.}<1998>]{maoz+etal}
{Maoz} D., {Koratkar} A., {Shields} J.~C., {Ho} L.~C., {Filippenko} A.~V.,
  {Sternberg } A., 1998,
\newblock AJ, {116}, 55

\bibitem[{Markert} \& {Donahue}<1985>]{markert}
{Markert} T.~H., {Donahue} M.~E., 1985,
\newblock ApJ, {297}, 564

\bibitem[{Markert} \& {Rallis}<1983>]{markert+rallis}
{Markert} T.~H., {Rallis} A.~D., 1983,
\newblock ApJ, {275}, 571

\bibitem[{Matonick} \& {Fesen}<1997>]{matonick2}
{Matonick} D.~M., {Fesen} R.~A., 1997,
\newblock ApJS, {112}, 49

\bibitem[{Matonick} {et~al.}<1997>]{matonick}
{Matonick} D.~M., {Fesen} R.~A., {Blair} W.~P., {Long} K.~S., 1997,
\newblock ApJS, {113}, 333

\bibitem[{McHardy} {et~al.}<1998>]{mchardy+etal}
{McHardy} I.~M., {Jones} L.~R., {Merrifield} M.~R., {Mason} K.~O., {Newsam}
  A.~M., {Abraham} R.~G., {Dalton} G.~B., {Carrera} F., {Smith} P.~J.,
  {Rowan-Robinson} M., {Abraham} R.~G., 1998,
\newblock mnras, {295}, 641

\bibitem[{McLeod} \& {Rieke}<1995>]{mcleod3}
{McLeod} K.~K., {Rieke} G.~H., 1995,
\newblock ApJ, {441}, 96

\bibitem[{McLeod}, {Rieke} \& {Storrie-Lombardi}<1999>]{mcleod2}
{McLeod} K.~K., {Rieke} G.~H., {Storrie-Lombardi} L.~J., 1999,
\newblock ApJ, {511}, L67

\bibitem[{McLeod}<1997>]{mcleod}
{McLeod} K.~K., 1997,
\newblock in Quasar Hosts, eds, {Clements} D.~L., {Perez-Fournon} I.

\bibitem[{McLure} {et~al.}<1999>]{mclure}
{McLure} R.~J., {Kukula} M.~J., {Dunlop} J.~S., {Baum} S.~A., {O'Dea} C.~P.,
  {Hughes} D.~H., 1999,
\newblock MNRAS, {308}, 377

\bibitem[{Moran}, {Halpern} \& {Helfand}<1996>]{moran}
{Moran} E.~C., {Halpern} J.~P., {Helfand} D.~J., 1996,
\newblock ApJS, {106}, 341

\bibitem[{Morris}<1993>]{morris}
{Morris} M., 1993,
\newblock ApJ, {408}, 496

\bibitem[{Papaderos} {et~al.}<1994>]{papaderos}
{Papaderos} P., {Fricke} K.~J., {Thuan} T.~X., {Loose} H.~H., 1994,
\newblock A\&A, {291}, L13

\bibitem[{Persic} {et~al.}<1998>]{persic}
{Persic} M., {Mariani} S., {Cappi} M., {Bassani} L., {Danese} L., {Dean} A.~J.,
  {Di Cocco} G., {Franceschini} A., {Hunt} L.~K., {Matteucci} F., {Palazzi} E.,
  {Palumbo} G. G.~C., {Rephaeli} Y., {Salucci} P., {Spizzichino} A., 1998,
\newblock A\&A, {339}, L33

\bibitem[{Petre} {et~al.}<1993>]{petre+etal}
{Petre} R., {Mushotzky} R.~F., {Serlemitsos} P.~J., {Jahoda} K., {Marshall}
  F.~E., 1993,
\newblock ApJ, {418}, 644+

\bibitem[{Primini}, {Forman} \& {Jones}<1993>]{primini}
{Primini} F.~A., {Forman} W., {Jones} C., 1993,
\newblock ApJ, {410}, 615

\bibitem[{Ptak} {et~al.}<1997>]{ptak+etal}
{Ptak} A., {Serlemitsos} P., {Yaqoob} T., {Mushotzky} R., {Tsuru} T., 1997,
\newblock AJ, {113}, 1286

\bibitem[{Ptak} {et~al.}<1999>]{ptak+etal2}
{Ptak} A., {Serlemitsos} P., {Yaqoob} T., {Mushotzky} R., 1999,
\newblock ApJS, {120}, 179

\bibitem[{Read}, {Ponman} \& {Strickland}<1997>]{read}
{Read} A.~M., {Ponman} T.~J., {Strickland} D.~K., 1997,
\newblock MNRAS, {286}, 626

\bibitem[{Roberts} \& {Warwick}<2000>]{roberts+warwick}
{Roberts} T.~P., {Warwick} R.~S.,
\newblock 2000,
\newblock MNRAS in press

\bibitem[{Roberts}, {Warwick} \& {Ohashi}<1999>]{roberts+etal}
{Roberts} T.~P., {Warwick} R.~S., {Ohashi} T., 1999,
\newblock MNRAS, {304}, 52

\bibitem[{Sadler}, {Jenkins} \& {Kotanyi}<1989>]{sadler}
{Sadler} E.~M., {Jenkins} C.~R., {Kotanyi} C.~G., 1989,
\newblock MNRAS, {240}, 591

\bibitem[{Sandage} \& {Tammann}<1981>]{sandage+tamman}
{Sandage} A., {Tammann} G.~A., 1981,
\newblock A Revised Shapley-Ames Catalog of Bright Galaxies (RSA)

\bibitem[{Schlegel}<1994a>]{schlegel1}
{Schlegel} E.~M., 1994a,
\newblock AJ, {108}, 1893

\bibitem[{Schlegel}<1994b>]{schlegel2}
{Schlegel} E.~M., 1994b,
\newblock ApJ, {424}, L99

\bibitem[{Schlegel}<1994c>]{schlegel3}
{Schlegel} E.~M., 1994c,
\newblock ApJ, {434}, 523

\bibitem[{Schmidt} {et~al.}<1998>]{schmidt+etal}
{Schmidt} M., {Hasinger} G., {Gunn} J., {Schneider} D., {Burg} R., {Giacconi}
  R., {Lehmann} I., {MacKenty} J., {Trumper} J., {Zamorani} G., 1998,
\newblock aa, {329}, 495

\bibitem[{Schulman} \& {Bregman}<1995>]{schulman+bregman}
{Schulman} E., {Bregman} J.~N., 1995,
\newblock ApJ, {441}, 568

\bibitem[Serlemitsos, Ptak \& Yaqoob<1996>]{serlemitsos+ptak+yaqoob}
Serlemitsos P., Ptak A., Yaqoob T., 1996,
\newblock in ASP Conf. Ser. 103, The physics of LINERs in view of recent
  observations, eds, {Eracleous} M., {Koratkar} A.~P., {Leitherer} C., {Ho} L.

\bibitem[{Simien} \& {De Vaucouleurs}<1986>]{simien}
{Simien} F., {De Vaucouleurs} G., 1986,
\newblock ApJ, {302}, 564

\bibitem[{Snowden} \& {Pietsch}<1995>]{snowden}
{Snowden} S.~L., {Pietsch} W., 1995,
\newblock ApJ, {452}, 627

\bibitem[{Stark} {et~al.}<1992>]{stark+etal}
{Stark} A.~A., {Gammie} C.~F., {Wilson} R.~W., {Bally} J., {Linke} R.~A.,
  {Heiles} C., {Hurwitz} M., 1992,
\newblock ApJS, {79}, 77

\bibitem[{Supper} {et~al.}<1997>]{supper}
{Supper} R., {Hasinger} G., {Pietsch} W., {Truemper} J., {Jain} A., {Magnier}
  E.~A., {Lewin} W. H.~G., {Van Paradijs} J., 1997,
\newblock A\&A, {317}, 328

\bibitem[{Takano} {et~al.}<1994>]{takano}
{Takano} M., {Mitsuda} K., {Fukazawa} Y., {Nagase} F., 1994,
\newblock ApJ, {436}, L47

\bibitem[{Trinchieri} \& {Fabbiano}<1991>]{trinchieri+fabbiano}
{Trinchieri} G., {Fabbiano} G., 1991,
\newblock ApJ, {382}, 82

\bibitem[{Trinchieri} {et~al.}<1999>]{trinchieri+etal3}
{Trinchieri} G., {Israel} G.~L., {Chiappetti} L., {Belloni} T., {Stella} L.,
  {Primini} F., {Fabbiano} P., {Pietsch} W., 1999,
\newblock A\&A, {348}, 43

\bibitem[{Trinchieri}, {Fabbiano} \& {Paulumbo}<1985>]{trinchieri+etal}
{Trinchieri} G., {Fabbiano} G., {Paulumbo} G. G.~C., 1985,
\newblock ApJ, {290}, 96

\bibitem[{Trinchieri}, {Fabbiano} \& {Romaine}<1990>]{trinchieri+etal2}
{Trinchieri} G., {Fabbiano} G., {Romaine} S., 1990,
\newblock ApJ, {356}, 110

\bibitem[{Van Der Marel}, {De Zeeuw} \& {Rix}<1997>]{vandermarel+etal}
{Van Der Marel} R.~P., {De Zeeuw} P.~T., {Rix} H.-W., 1997,
\newblock ApJ, {488}, 119

\bibitem[{Van Der Marel}<1999>]{vandermarel1}
{Van Der Marel} R.~P., 1999,
\newblock AJ, {117}, 744

\bibitem[{Van Dyk}, {Hamuy} \& {Filippenko}<1996>]{vandyk}
{Van Dyk} S.~D., {Hamuy} M., {Filippenko} A.~V., 1996,
\newblock AJ, {111}, 2017

\bibitem[{Van Paradijs} \& {McClintock}<1995>]{vanparadijs+mcclintock}
{Van Paradijs} J., {McClintock} J., 1995,
\newblock in X-ray binaries - Cambridge Astrophysics Series, Cambridge, MA:
  Cambridge University Press, eds, {Lewin} W.~H., {Van Paradijs} J., {Van Den
  Heuvel} E.~P.

\bibitem[{Vogler} \& {Pietsch}<1999>]{vogler+pietsch}
{Vogler} A., {Pietsch} W., 1999,
\newblock A\&A, {342}, 101

\bibitem[{Wang}, {Immler} \& {Pietsch}<1999>]{wang+etal}
{Wang} D., {Immler} S., {Pietsch} W.,
\newblock 1999,
\newblock preprint, astro-ph/9903479

\bibitem[{Wang}<1999>]{wang}
{Wang} D.,
\newblock 1999,
\newblock preprint, astro-ph/9903246

\bibitem[{Williams} \& {Chu}<1995>]{murphy}
{Williams} R.~M., {Chu} Y.-H., 1995,
\newblock ApJ, {439}, 132

\bibitem[{Yee}<1992>]{yee.92}
{Yee} H.~K., 1992,
\newblock in ASP Conf. Ser. 31, Relationships Between Active Galactic Nuclei
  and Starburst Galaxies, ed., Filippenko A.~V.

\end{thebibliography}

\end{document}